\documentclass[pra,twocolumn,superscriptaddress,showpacs,nofootinbibfloatfix,amsmath,amsfonts,amssymb]{revtex4-1}%
\usepackage{amsmath,amsfonts,amssymb,color}
\usepackage{amsthm}
\usepackage{leftidx}
\usepackage{graphicx}
\usepackage{xcolor}
\usepackage{dcolumn}
\usepackage{bm}
\usepackage{epstopdf}
\usepackage{epsfig}
\usepackage{environ}
\usepackage{pdfcomment}

\usepackage{float}
\usepackage[T1]{fontenc}
\usepackage[latin9]{inputenc}
\usepackage{setspace}
\usepackage{esint}

\begin{document}

\preprint{APS/123-QED}

\title{\textbf{Non-Hermitian Floquet topological superconductors with multiple Majorana edge modes}}

\author{Longwen Zhou}
\email{zhoulw13@u.nus.edu}
\affiliation{%
	Department of Physics, College of Information Science and Engineering, Ocean University of China, Qingdao 266100, China
}
\affiliation{Institute of Theoretical Physics, Chinese Academy of Sciences, Beijing 100190, China
}

\date{\today}

\begin{abstract}
Majorana edge modes are candidate elements of topological quantum
computing. In this work, we propose a Floquet engineering approach
to generate multiple non-Hermitian Majorana zero and $\pi$
modes at the edges of a one-dimensional topological superconductor.
Focusing on a Kitaev chain with periodically kicked superconducting
pairings and gain/losses in the chemical potential or nearest
neighbor hopping terms, we found rich non-Hermitian Floquet topological
superconducting phases, which are originated from the interplay between
drivings and non-Hermitian effects. Each of the phases is characterized
by a pair of topological winding numbers, which can in principle take
arbitrarily large integer values thanks to the applied driving fields.
Under the open boundary condition, these winding numbers also predict
the number of degenerate Majorana edge modes with quasienergies zero
and $\pi$. Our findings thus expand the family of Floquet topological
phases in non-Hermitian settings, with potential applications in realizing
environmentally robust Floquet topological quantum computations.
\end{abstract}

\maketitle

\section{Introduction}\label{sec:Int}

Non-Hermitian topological states of matter have attracted great attention in recent years~\cite{NHBook1,NHBook2,NHTPReview1,NHTPReview2,NHTPReview3,NHTPReview4,NHTPReview5,NHTPReview6,NHTPReview7,NHTPReview8,NHTPReview9,NHTPReview10}. These exotic phases could appear in open systems with gain/loss~\cite{NHTPOS1,NHTPOS2,NHTPOS3,NHTPOS4}, bosonic systems with dynamical instability~\cite{NHTPBOSE1,NHTPBOSE2,NHTPBOSE3,NHTPBOSE4,NHTPBOSE5,NHTPBOSE6,NHTPBOSE7}, electronic systems with finite-lifetime quasiparticles~\cite{NHTPFERMI1,NHTPFERMI2,NHTPFERMI3,NHTPFERMI4,NHTPFERMI5,NHTPFERMI6,NHTPFERMI7} and mechanical metamaterials~\cite{NHTPMETA1,NHTPMETA2,NHTPMETA3}, with potential applications in achieving unidirectional invisibility~\cite{NHTPInvis1,NHTPInvis2,NHTPInvis3,NHTPInvis4}, enhanced sensitivity~\cite{NHTPSens1,NHTPSens2,NHTPSens3,NHTPSens4,NHTPSens5} and topological lasers~\cite{TPLaser1,TPLaser2,TPLaser3,TPLaser4,TPLaser5,TPLaser6,TPLaser7,TPLaser8}. Theoretically, various schemes of classifying non-Hermitian topological phases in the absence or presence of interactions have been proposed~\cite{Class1,Class2,Class3,Class4,Class5,Class6,Class7,Class8,Class9,Class10,Class11,Class12,Class13}, and the related bulk-boundary correspondence has been investigated in detail~\cite{BBC0,BBC1,BBC2,BBC3,BBC4,BBC5,BBC6,BBC7,BBC8,BBC9,BBC10,BBC11,BBC12}. Experimentally, non-Hermitian topological states of matter have been observed in mechanical~\cite{Exp51,Exp52}, optical~\cite{Exp11,Exp12,Exp13}, photonic~\cite{Exp21,Exp22,Exp23}, optomechanical~\cite{Exp41,Exp42} systems and topolectric circuits~\cite{Exp31,Exp32}. Recently, non-Hermitian topological phases with large topological invariants~\cite{ZhouPRB2019,ZhouPRB2018,ZhouPRA2019}, many edge states~\cite{ZhouPRB2018,ZhouPRA2019,FehskePRL2019,LiPRB2019,YuceEPJD2015}, non-Hermitian skin effects~\cite{NHSkin12} and quantized pumps~\cite{NHPump1} have also been found in periodically driven (Floquet) systems.

In applications, non-Hermitian topological states of matter in superconducting
systems are one of the most intriguing candidates, as they may possess
degenerate Majorana modes at their boundaries. The Majorana edge modes
have been investigated intensively in the past decade due to their
potential applications in topological quantum computing (see
\cite{MajoranaRev3,MajoranaRev4,MajoranaRev5,MajoranaRev6,MajoranaRev7,MajoranaRev8,MajoranaRev9} for reviews). The emergence and persistence of Majorana edge modes
in non-Hermitian settings may further provide evidence for their
robustness to environmental effects and certain other effects like the quasiparticle poisoning. Recently, several studies have
touched upon the topic of Majorana modes in non-Hermitian systems~\cite{NHMajorana0,NHMajorana1,NHMajorana2,NHMajorana3,NHMajorana4,NHMajorana5,NHMajorana6,NHMajorana7,NHMajorana8,NHMajorana9,NHMajorana11,NHMajorana12,NHMajorana13,NHMajorana14},
which possess unique phenomena like the coalescing~\cite{NHMajorana4} due to the presence of exceptional points and the nonlocal particle transport~\cite{NHMajorana2} due to the interplay between non-Hermiticity and superconductivity. It was also
suggested that putting topological superconductors in the non-Hermitian
setting could be helpful to resolve the controversy between Andreev
and Majorana bound states~\cite{NHMajorana9}.

However, most of the above mentioned studies focus on the generation
and characterization of Majorana edge modes in static non-Hermitian
systems. In periodically driven (Floquet) systems, due to the long-range
hopping induced by the driving fields, superconducting phases characterized
by large topological invariants and multiple topological edge states
could appear~\cite{FloMaj0,FloMaj1,FloMaj2,FloMaj3,FloMaj4,FloMaj5,FloMaj6,FloMaj7,FloMaj8}. 
In one-dimension (1d), the resulting Floquet Majorana
edge modes have been further employed in a recent proposal of Floquet
topological quantum computation~\cite{FloCompt1,FloCompt2,FloCompt3,FloCompt4}. 
Nevertheless, in actual topological superconductors, the Majorana modes could subject to quasiparticle poisoning effects and various types of decoherence effects induced by the environment. As these effects might be captured by adding non-Hermitian terms to the system's Hamiltonian, it is interesting to check whether the multiple Majorana edge modes generated by Floquet engineering could still persist in such non-Hermitian settings. Furthermore, the periodic driving and non-Hermitian effects may cooperate to create new topological superconducting phases that can appear neither in closed Floquet systems nor in static non-Hermitian systems.

Motivated by the above considerations, we extend the Floquet
engineering approach to $1$d non-Hermitian topological superconducting
systems in this work,
with a threefold purpose.
First, we plan to study how non-Hermiticity can influence the topological properties of periodically driven systems.
Second, we plan to find topological superconducting phases
with large topological invariants that are unique to non-Hermitian Floquet systems. Third, we would like to obtain multiple Majorana zero and $\pi$ edge modes even when there is dissipation, which might allow them to be useful in the situations with environmental effects, or to be able to combat certain quasiparticle poisoning effects.
As the Kitaev chain (KC)~\cite{KC} is the paradigmatic model in the study
of 1d topological superconductors and Majorana edge modes, we start
with a brief review of the non-Hermitian extension of KC.
After that, we introduce our Floquet version of the non-Hermitian
KC (NHKC) via periodically kicking the superconducting pairing
terms. In Sec.~\ref{sec:PKNHKCBulk}, we investigate the bulk properties
of the periodically kicked NHKC systematically, focusing on the symmetries
that protect its topological phases and the invariants that characterize
its topological states. We establish the bulk topological phase diagram
of the periodically kicked NHKC, where rich non-Hermitian Floquet
topological superconducting phases are found. Furthermore, we show
in Sec.~\ref{sec:Majorana} the Floquet Majorana edge modes with real quasienergies
zero and $\pi$ in the system under the open boundary condition~(OBC), and
relate their numbers to the bulk topological invariants of the model. 
In Sec.~\ref{sec:Exp}, we discuss experimental issues regarding the realization of our model and the detection of its topological properties.
In Sec.~\ref{sec:Summary}, we summarize our results and discuss potential
future directions.

\section{Model and symmetry\label{sec:KCM}}

In this section, we first introduce the non-Hermitian variant of KC and discuss its known physical properties. This is followed by a presentation of our periodically driven NHKC model in both lattice and momentum representations. The non-Hermitian Floquet topological phases and Majorana edge modes that can appear in this model are the focus of our study in the later parts of this work.

\subsection{Non-Hermitian Kitaev chain\label{subsec:NHKC}}

In this subsection, we briefly review the non-Hermitian extension of KC model, with a focus on its symmetry and topological properties. The KC provides a paradigmatic example for the study of topological superconductors and Majorana edge modes~\cite{KC}. It describes spinless electrons in a 1d tight-binding lattice with $p$-wave pairings. The model Hamiltonian is given by
\begin{alignat}{1}
H= & \frac{1}{2}\sum_{n}[\mu(2c_{n}^{\dagger}c_{n}-1)+J(c_{n}^{\dagger}c_{n+1}+{\rm H.c.})]\nonumber \\
+ & \frac{1}{2}\sum_{n}\Delta(e^{i\phi}c_{n}^{\dagger}c_{n+1}^{\dagger}+{\rm H.c.}),\label{eq:KC}
\end{alignat}
where $c_{n}^{\dagger}$ ($c_{n}$) is the creation (annihilation) operator
of an electron on the lattice site $n$, $\mu$ is the chemical potential
and $J$ is the nearest neighbor hopping amplitude. $\Delta\in\mathbb{R}$
characterizes the strength of $p$-wave superconducting pairing and
$\phi$ is the superconducting pairing phase, which can in principle take an arbitrary value in $[0,2\pi)$.

For a chain of length $L$ and under the periodic boundary condition, the Hamiltonian
$H$ in Eq.~(\ref{eq:KC}) can be expressed in the momentum representation in terms of
Fourier transforms $c_{n}=\frac{1}{\sqrt{L}}\sum_{k}c_{k}e^{ikn}$
and $c_{n}^{\dagger}=\frac{1}{\sqrt{L}}\sum_{k}c_{k}^{\dagger}e^{-ikn}$,
yielding 
\begin{equation}
H=\frac{1}{2}\sum_{k\in{\rm BZ}}\Psi_{k}^{\dagger}H_{\phi}(k)\Psi_{k}.
\end{equation}
Here $k\in[-\pi,\pi)$ is the quasimomentum, $\Psi_{k}^{\dagger}\equiv(c_{k}^{\dagger},c_{-k})$
is the Nambu spinor, and the Hamiltonian matrix
\begin{equation}
H_{\phi}(k)=(\mu+J\cos k)\sigma_{z}-\Delta\sin(k)[\sin(\phi)\sigma_{x}+\cos(\phi)\sigma_{y}]
\end{equation}
where $\sigma_{x,y,z}$ are Pauli matrices in their usual representations.
The dispersion relation of the system $E_{\pm}(k)=\pm E(k)$ can
further be obtained by performing the Bogoliubov transformation~\cite{Book3}, yielding
\begin{equation}
E(k)=\sqrt{\Delta^{2}\sin^{2}(k)+(\mu+J\cos k)^{2}}.
\end{equation}
Note in passing that the excitation energy $E(k)$ is independent
of the pairing phase $\phi$. From now on, we choose the pairing phase
$\phi=\pi$, which leads to the most frequently studied version of
KC. The Hamiltonian matrix $H_\phi(k)$ then simplifies
to
\begin{equation}
H_{\pi}(k)\equiv h_{y}(k)\sigma_{y}+h_{z}(k)\sigma_{z},\label{eq:Hpik}
\end{equation}
where the two components in front of Pauli matrices $\sigma_{y,z}$ are
\begin{equation}
h_{y}(k)=\Delta\sin(k),\qquad h_{z}(k)=\mu+J\cos k.\label{eq:hxyk}
\end{equation}
It is clear that the Hamiltonian matrix $H_{\pi}(k)$ of KC
possesses the sublattice symmetry $\Gamma=\sigma_{x}$, in the sense
that
\begin{equation}
\Gamma H_{\pi}(k)\Gamma=-H_{\pi}(k),\qquad\Gamma^{2}=\sigma_{0},
\end{equation}
where $\sigma_{0}$ is the $2\times2$ identity matrix. When $\mu$
and $J$ take real values, $H_{\pi}(k)$ also possesses the time-reversal
symmetry ${\cal T}=\sigma_{0}$ and particle-hole symmetry ${\cal C}=\sigma_{x}$,
in the sense that
\begin{alignat}{1}
\sigma_{0}H_{\pi}^{*}(k)\sigma_{0}= & H_{\pi}(-k),\label{eq:HpiTRS}\\
\sigma_{x}H_{\pi}^{*}(k)\sigma_{x}= & -H_{\pi}(-k),\label{eq:HpiPHS}
\end{alignat}
with $\sigma_{0}\sigma_{0}^{*}=\sigma_{0}$ and $\sigma_{x}\sigma_{x}^{*}=\sigma_{0}$,
respectively. The Hamiltonian $H_{\pi}(k)$ then belongs to the symmetry
class BDI~\cite{Tenfold}. Each of its superconducting phases is characterized by
an integer topological winding number $w$, defined as
\begin{equation}
w=\int_{-\pi}^{\pi}\frac{dk}{2\pi}\frac{h_{z}(k)\partial_{k}h_{y}(k)-h_{y}(k)\partial_{k}h_{z}(k)}{h_{y}^{2}(k)+h_{z}^{2}(k)}.\label{eq:W}
\end{equation}
Under the OBC, the number of Majorana zero modes
localized at each boundary of the chain is equal to $|w|$, which
reflects the general principle of bulk-edge correspondence. The degeneracy of these Majorana modes is protected by the
particle-hole symmetry $\sigma_{x}$.

Even though the original version of KC~\cite{KC} belongs to the symmetry class D with an arbitrary pairing phase $\phi$ due to the broken $U(1)$ symmetry, the KC described by $H_\pi(k)$ in the symmetry class BDI has also been thoroughly explored in the context of solid state physics~(see \cite{MajoranaRev7,MajoranaRev8,MajoranaRev9} for reviews). It is also relevant to the realization of topological superfluids and superconductors in cold atom systems~\cite{CooperRMPCold}. In the meantime, the KC in BDI class has been considered by several studies in the context of Floquet engineering~\cite{FloMaj3,FloMaj4,FloMaj8}. Compared with the original KC~\cite{KC} in symmetry class D, one of the most attractive features of the KC in BDI class is that it is characterized by an integer~(${\mathbb Z}$) topological invariant. This means that it allows more than one pair of Majorana edge modes to appear under the OBC, which is beneficial for the realization of topological quantum computing via braiding Majorana modes. Furthermore, it has also been shown that the interactions between fermions preserving time-reversal symmetry can change the ${\mathbb Z}$ classification of 1d models in BDI class to the ${\mathbb Z}_8$ classification, gapping out eight Majorana modes~\cite{KCInt}. Therefore, the KC in BDI class is also relevant for the study of symmetry-protected topological phases in 1d with many-body interactions.

An NHKC can now be obtained by setting the hopping
amplitude $J$ or the chemical potential $\mu$ to complex values,
yielding $H_{\pi}^{\dagger}(k)\neq H_{\pi}(k)$ in Eq.~(\ref{eq:Hpik}).
Physically, these choices may be realized by introducing lossy
effects to the nearest neighbor hopping amplitude ($J\in\mathbb{C}$),
or adding onsite particle gain/losses ($\mu\in\mathbb{C}$). 
Experimentally, the NHKC may be realized by loading fermionic cold atoms in a 1d optical lattice, where the effective $p$-wave pairing can be induced by an optical Raman transition~\cite{FloMaj0}, and the non-Hermiticity may be implemented by controlling and monitoring the decay of atoms~\cite{NHKCExp1,NHKCExp2,NHKCExp3}.
However, in these cases the time-reversal and particle-hole symmetries as defined
in Eqs.~(\ref{eq:HpiTRS}) and (\ref{eq:HpiPHS}) are not hold. A
simple reason behind this anomaly is that while the transpose ($\top$)
and complex conjugate ($*$) of a Hermitian matrix are the same, they
are different for a non-Hermitian matrix, i.e., $H_{\pi}^{\dagger}(k)\neq H_{\pi}(k)\Leftrightarrow H_{\pi}^{\top}(k)\neq H_{\pi}^{*}(k)$.
Recently, progresses have been made in the symmetry and topological
classification of non-Hermitian matrices. In Ref.~\cite{Class7}, a classification
scheme for non-Hermitian topological phases is introduced by generalizing
the ``tenfold way'' of topological insulators and superconductors~\cite{Tenfold}
to a $38$-fold ``periodic table''. In this enlarged periodic table,
the matrix transposition and complex conjugation are treated as independent
operations, each generating its own flavor of ``time-reversal'' and ``particle-hole''
symmetries. According to this new classification scheme, the KC Hamiltonian $H_{\pi}(k)$ in Eq.~(\ref{eq:Hpik}), with $J\in\mathbb{C}$
or $\mu\in\mathbb{C}$, belongs to a generalized BDI class. It possesses
the modified version of time-reversal (${\cal T}=\sigma_{0}$) and
particle-hole (${\cal C}=\sigma_{x}$) symmetries, defined with respect to the matrix transposition as~\cite{Class7,NHMajorana1}
\begin{alignat}{1}
\sigma_{0}H_{\pi}^{\top}(k)\sigma_{0}= & H_{\pi}(-k),\label{eq:HpiTRST}\\
\sigma_{x}H_{\pi}^{\top}(k)\sigma_{x}= & -H_{\pi}(-k).\label{eq:HpiPHST}
\end{alignat}
The topological phases of the NHKC can still be characterized by the winding number $w$ in Eq.~(\ref{eq:W}) under the periodic boundary condition.

In recent studies, several variants of the NHKC have been investigated in detail~\cite{NHMajorana0,NHMajorana1,NHMajorana2,NHMajorana3,NHMajorana4,NHMajorana5,NHMajorana6,NHMajorana7,NHMajorana8,NHMajorana9,NHMajorana11,NHMajorana12,NHMajorana13,NHMajorana14}. Some topological phases unique to
non-Hermitian superconducting systems were identified, and degenerate
Majorana zero modes were found numerically in the KC with
losses. Potential applications of these findings include the realization
of topological quantum computing, and the resolution of the controversy
between Andreev and Majorana modes~\cite{NHMajorana9}. However, most of these studies
are limited to static non-Hermitian systems carrying a limited number
of Majorana edge modes. This motivates us to introduce the Floquet
extension of the NHKC in the following subsection,
which will be shown to possess an enriched topological phase diagram together
with a large number of coexisting Majorana zero and $\pi$ edge modes.

\subsection{Periodically kicked non-Hermitian Kitaev chain}

In this subsection, we introduce our Floquet extension of the NHKC, whose topological properties will be investigated in detail
in later sections. Our model describes a KC, whose superconducting
pairing term is kicked periodically by a delta-shaped pulse, with the
pairing phase fixed at $\phi=\pi$. The model Hamiltonian then takes
the form
\begin{alignat}{1}
H(t)= & \frac{1}{2}\sum_{n}[\mu(2c_{n}^{\dagger}c_{n}-1)+J(c_{n}^{\dagger}c_{n+1}+{\rm H.c.})],\nonumber \\
- & \frac{1}{2}\sum_{n}\Delta\delta_{T}(t)(c_{n}^{\dagger}c_{n+1}^{\dagger}+{\rm H.c.}),
\end{alignat}
where $\delta_{T}(t)\equiv\sum_{\ell\in\mathbb{Z}}\delta(t/T-\ell)$
describes delta kickings with the period $T$, and the system
is made non-Hermitian by assuming $J\in\mathbb{C}$ or $\mu\in\mathbb{C}$.
Under the periodic boundary condition, the Hamiltonian $H(t)$ can
be expressed in momentum space as $H(t)=\frac{1}{2}\sum_{k\in{\rm BZ}}\Psi_{k}^{\dagger}H(k,t)\Psi_{k}$,
where $\Psi_{k}$ is the same Nambu spinor operator as defined in
the last subsection, and 
\begin{equation}
H(k,t)=h_{y}(k)\delta_{T}(t)\sigma_{y}+h_{z}(k)\sigma_{z}.\label{eq:Hkt}
\end{equation}
Here $h_{y,z}(k)$ are given explicitly by Eq.~(\ref{eq:hxyk}).
The Floquet operator of the system, which describes its time evolution
from $t=\ell T+0^{-}$ to $(\ell+1)T+0^{-}$ with $\ell\in{\mathbb Z}$ is $U(k)=e^{-i\frac{T}{\hbar}h_{z}(k)\sigma_{z}}e^{-i\frac{T}{\hbar}h_{y}(k)\sigma_{y}}$
[See Appendix~\ref{subsec:Uk} for more details regarding how to arrive at $U(k)$ from Eq.~(\ref{eq:Hkt})].
Choosing the unit of energy to be $\hbar/T$ and setting $\hbar=T=1$,
we can express the Floquet operator in dimensionless units as
\begin{equation}
U(k)=e^{-ih_{z}(k)\sigma_{z}}e^{-ih_{y}(k)\sigma_{y}}.\label{eq:UkPKNHKC}
\end{equation}
By expanding each exponential term in $U(k)$ with the Euler formula,
we can further obtain the dispersion relation of Floquet quasienergy
bands as $\varepsilon_{\pm}(k)=\pm\varepsilon(k)$, with
\begin{equation}
\varepsilon(k)=\arccos\left\{ \cos[h_{y}(k)]\cos[h_{z}(k)]\right\} .\label{eq:varepk}
\end{equation}
Note that $\varepsilon(k)$ could become complex if the chemical potential
$\mu$ or hopping amplitude $J$ in $h_{z}(k)$ take complex values. However, only the real part of $\varepsilon(k)$ corresponds to a phase factor, which could have periodicity in the quasienergy Brillouin zone.
In the following section, we will analyze in detail the bulk topological
properties of this periodically kicked NHKC.

\section{Topological invariants and phase diagram\label{sec:PKNHKCBulk}}

The periodically kicked NHKC introduced in the last section
possesses rich non-Hermitian Floquet topological superconducting phases.
In this section, we first reveal the underlying symmetries of $U(k)$
and the winding numbers that characterize its topological phases.
The bulk phase diagrams of the periodically kicked NHKC are then presented in some typical parameter regions, with each of the phases being characterized by the introduced winding numbers.

\subsection{Symmetry and topological winding numbers}

The topological phases that can appear in the Floquet system described by
$U(k)$ in Eq.~(\ref{eq:UkPKNHKC}) depend on the symmetries that $U(k)$ possesses. Following
earlier studies on the topological classification of Floquet systems~\cite{AsbothSTF},
we first express $U(k)$ in a pair of symmetric time frames as
\begin{alignat}{1}
U_{1}(k)= & e^{-i\frac{h_{y}(k)}{2}\sigma_{y}}e^{-ih_{z}(k)\sigma_{z}}e^{-i\frac{h_{y}(k)}{2}\sigma_{y}},\\
U_{2}(k)= & e^{-i\frac{h_{z}(k)}{2}\sigma_{z}}e^{-ih_{y}(k)\sigma_{y}}e^{-i\frac{h_{z}(k)}{2}\sigma_{z}}.
\end{alignat}
It is clear that the Floquet operators $U(k)$, $U_{1}(k)$ and $U_{2}(k)$
are related by $k$-dependent similarity transformations. Therefore,
they share the same quasienergy spectrum $\pm\varepsilon(k)$ as given
by Eq.~(\ref{eq:varepk}), but may possess different topological properties.
Moreover, employing again the Euler formula, we can express $U_{1}(k)$
and $U_{2}(k)$ in terms of effective Floquet Hamiltonians as
\begin{equation}
U_{1}(k)=e^{-iH_{1}(k)},\qquad U_{2}(k)=e^{-iH_{2}(k)}.\label{eq:U12k}
\end{equation}
The effective Hamiltonians $H_{1}(k)$ and $H_{2}(k)$ take the
form
\begin{equation}
H_{\alpha}(k)=h_{\alpha y}(k)\sigma_{y}+h_{\alpha z}(k)\sigma_{z}\label{eq:H12k}
\end{equation}
for $\alpha=1,2$, where the components $h_{\alpha y}(k)$ and $h_{\alpha z}(k)$
are explicitly given by:
\begin{alignat}{1}
h_{1y}(k)= & \varepsilon(k)\sin[h_{y}(k)]\cos[h_{z}(k)]/\sin[\varepsilon(k)],\\
h_{1z}(k)= & \varepsilon(k)\sin[h_{z}(k)]/\sin[\varepsilon(k)],\\
h_{2y}(k)= & \varepsilon(k)\sin[h_{y}(k)]/\sin[\varepsilon(k)],\\
h_{2z}(k)= & \varepsilon(k)\cos[h_{y}(k)]\sin[h_{z}(k)]/\sin[\varepsilon(k)].
\end{alignat}
According to Eqs.~(\ref{eq:hxyk}) and (\ref{eq:varepk}), we have
$h_{y}(-k)=-h_{y}(k)$, $h_{z}(-k)=h_{z}(k)$ and $\varepsilon(-k)=\varepsilon(k)$.
Therefore, following the discussions in Subsec.~\ref{subsec:NHKC},
both the effective Hamiltonians $H_{1}(k)$ and $H_{2}(k)$ possess
the generalized time-reversal symmetry ${\cal T}=\sigma_{0}$, particle-hole symmetry ${\cal C}=\sigma_{x}$ and sublattice symmetry $\Gamma=\sigma_{x}$ in the non-Hermitian setting, i.e.,
\begin{alignat}{1}
\sigma_{0}H_{\alpha}^{\top}(k)\sigma_{0}= & H_{\alpha}(-k),\label{eq:HaTRST}\\
\sigma_{x}H_{\alpha}^{\top}(k)\sigma_{x}= & -H_{\alpha}(-k),\label{eq:HaPHST}\\
\sigma_{x}H_{\alpha}(k)\sigma_{x}= & -H_{\alpha}(k),\label{eq:HaCST}
\end{alignat}
where $\alpha=1,2$. $H_{1}(k)$ and $H_{2}(k)$ then belong to the non-Hermitian extension of symmetry class BDI, with their topological phases being characterized by the winding number $w$ in Eq.~(\ref{eq:W}), i.e., 
\begin{equation}
w_{\alpha}=\int_{-\pi}^{\pi}\frac{dk}{2\pi}\frac{h_{\alpha z}(k)\partial_{k}h_{\alpha y}(k)-h_{\alpha y}(k)\partial_{k}h_{\alpha z}(k)}{h_{\alpha y}^{2}(k)+h_{\alpha z}^{2}(k)}\label{eq:W12}
\end{equation}
for $\alpha=1,2$. The topological phases of Floquet operator $U(k)$
in Eq.~(\ref{eq:UkPKNHKC}) then admit a $\mathbb{Z}\times\mathbb{Z}$
topological characterization. Namely, each of its Floquet topological phases
is characterized by a pair of winding numbers $(w_{0},w_{\pi})$, defined as
\begin{equation}
w_{0}\equiv\frac{w_{1}+w_{2}}{2}\qquad w_{\pi}\equiv\frac{w_{1}-w_{2}}{2}.\label{eq:W0P}
\end{equation}
Note that in the non-Hermitian regime, both $h_{\alpha y}(k)$ and $h_{\alpha z}(k)$ in Eq.~(\ref{eq:W12}) can in general be complex. However, what the Eq.~(\ref{eq:W12}) counts is the net number of times that the vectors Re$[h_{\alpha y}(k),h_{\alpha z}(k)]$ and Im$[h_{\alpha y}(k),h_{\alpha z}(k)]$ wind around their own zeros. The finite imaginary part of $[h_{\alpha y}(k),h_{\alpha z}(k)]$ would have no windings when tracing over the whole Brillouin zone~(see Appendix~\ref{subsec:ReW12} for a proof of this statement), and therefore it has no contribution to the topological winding number $w_\alpha$. These observations are also supported by our numerical calculations of the winding numbers in the next subsection.

In the following, we will demonstrate that these topological winding
numbers indeed characterize all the possible Floquet topological phases of
the periodically kicked NHKC. Moreover, in the
next section we will see that the values of $w_{0}$ and $w_{\pi}$
determine the number of degenerate Floquet Majorana edge modes at
quasienergies zero and $\pi$ under the OBC, respectively.

\subsection{Topological phase diagram of the periodically kicked NHKC}

In this subsection, we present the topological phase diagrams of the
periodically kicked NHKC. As shown in Eq.~(\ref{eq:varepk}), the
system possesses two quasienergy bands $\pm\varepsilon(k)$. The
real part of quasienergy $\varepsilon(k)$ is a phase factor defined
modulus $2\pi$, whereas the imaginary part of $\varepsilon(k)$ has no periodicity on the complex quasienergy plane. Therefore, the periodically kicked NHKC in general
holds two point-like band gaps around quasienergies $0$ and $\pm\pi$. The Floquet
spectrum of the system would become gapless when $\varepsilon(k)=0$ or
$\varepsilon(k)=\pi$~(i.e., closing a point gap). Assuming the chemical potential $\mu\in\mathbb{C}$
and hopping amplitude $J\in\mathbb{R}$, our model describes a periodically
kicked KC with onsite gain/loss. Its quasienergy spectrum becomes
gapless under the condition
\begin{equation}
\mu_{i}=\pm{\rm arccosh}\left\{ \frac{1}{\left|\cos\left[\Delta\sqrt{1-(n\pi-\mu_{r})^{2}/J^{2}}\right]\right|}\right\} ,\label{eq:Gapless1}
\end{equation}
where $n\in\mathbb{Z}$, $|n\pi-\mu_{r}|\leq|J|$, and $\mu_{r}$ ($\mu_{i}$)
is the real (imaginary) part of the chemical potential $\mu$ (see
Appendix \ref{subsec:GaplessCond} for the derivation details). Similarly,
when the hopping amplitude $J\in\mathbb{C}$ and chemical potential
$\mu\in\mathbb{R}$ our model describes a periodically kicked KC with
non-Hermitian hopping amplitudes. Its quasienergy spectrum would be
gapless under the condition
\begin{equation}
J_{i}=\pm\frac{J_{r}}{n\pi-\mu}{\rm arccosh}\left\{\frac{1}{\left|\cos\left[\Delta\sqrt{1-(n\pi-\mu)^{2}/J_{r}^{2}}\right]\right|}\right\},\label{eq:Gapless2}
\end{equation}
where $n\in\mathbb{Z}$, $|n\pi-\mu|\leq|J_{r}|$, and $J_{r}$ ($J_{i}$)
is the real (imaginary) part of the hopping amplitude $J$ (see Appendix
\ref{subsec:GaplessCond} for the derivation details). Eqs.~(\ref{eq:Gapless1})
and (\ref{eq:Gapless2}) determine the point-gap closing points
in the parameter space for two typical situations of our interest
(i.e., non-Hermiticity due to onsite gain/loss or non-Hermitian hopping).
The collection of these points are expected to form the boundaries between different
Floquet topological superconducting phases of the periodically kicked
NHKC. Moreover, each of the non-Hermitian Floquet topological phases
is characterized by the topological winding numbers $(w_{0},w_{\pi})$
in Eq.~(\ref{eq:W0P}).

\begin{figure}
	\begin{centering}
		\includegraphics[scale=0.49]{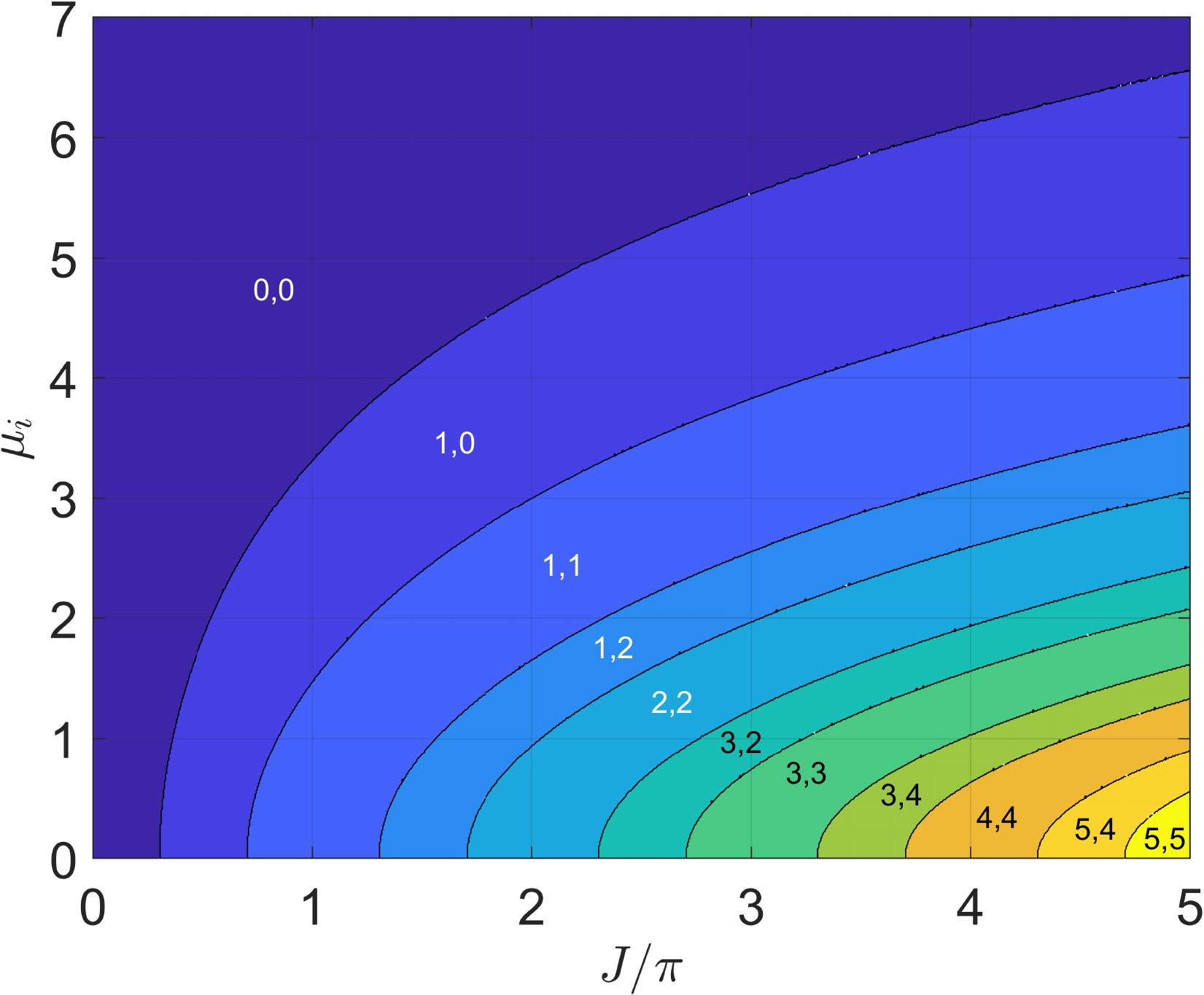}
		\par\end{centering}
	\caption{The topological phase diagram of the periodically kicked NHKC versus the
		hopping amplitude $J$ and the imaginary part of onsite chemical potential
		$\mu_{i}$. The real part of chemical potential and the superconducting
		pairing strength are set as $\mu_{r}=0.3\pi$ and $\Delta=0.5\pi$.
		Every patch with a uniform color corresponds to a non-Hermitian Floquet
		topological superconducting phase, characterized by the winding numbers
		$(w_{0},w_{\pi})$ as denoted on the patch. The black solid lines
		separating different patches are the topological phase boundaries
		determined by Eq.~(\ref{eq:Gapless1}).\label{fig:PhsDigm1}}
\end{figure}
In Fig.~\ref{fig:PhsDigm1}, we present the bulk topological phase
diagram of the periodically kicked NHKC with respect to the (real)
hopping amplitude $J$ and the imaginary part of the chemical potential
$\mu_{i}$. The real part of chemical potential and the superconducting
pairing strength are fixed at $\mu_{r}=0.3\pi$ and $\Delta=0.5\pi$,
respectively. In Eq.~(\ref{eq:Gapless1}), we also notice that the
phase boundaries possess reflection symmetries with respect to the
axis $\mu_{i}=0$ and $J=0$. Therefore, we only show the phase diagram
in a regime with $\mu_{i},J>0$. In Fig.~\ref{fig:PhsDigm1}, each region with a uniform
color corresponds to a non-Hermitian Floquet topological phase, characterized
by the winding numbers $(w_{0},w_{\pi})$ shown explicitly therein.
The black solid lines denote the boundaries between different topological
phases, which are calculated analytically from the point gap closing condition
Eq.~(\ref{eq:Gapless1}). We can see from this ``rainbow-shaped''
phase diagram that at each imaginary chemical potential $\mu_{i}$,
new Floquet topological superconducting phases emerge progressively
with the increase of the hopping amplitude $J$, which are characterized by larger
and larger topological winding numbers $(w_{0},w_{\pi})$. This tendency
happens to have no ends, and we could in principle obtain a phase
with arbitrarily large winding numbers $(w_{0},w_{\pi})$ by simply
increasing the strength of the nearest neighbor hopping amplitude $J$
with other system parameters $\Delta$ and $\mu$ fixed. Furthermore,
at a given hopping amplitude $J$, a series of topological phase transitions
can be induced by changing the imaginary part of chemical potential
$\mu_{i}$. These transitions are unique to non-Hermitian Floquet
systems. Each of them is accompanied by a quantized jump of the winding
numbers $(w_{0},w_{\pi})$, and the phase boundaries match perfectly
with our theoretical predictions in Eq.~(\ref{eq:Gapless1}). These
observations reveal the richness of Floquet topological phases in
the periodically kicked NHKC. The phases with large winding numbers
$(w_{0},w_{\pi})$ also imply the appearance of many Majorana zero
and $\pi$ edge modes in the system under the OBC,
as will be demonstrated in the next section.

\begin{figure}
	\begin{centering}
		\includegraphics[scale=0.49]{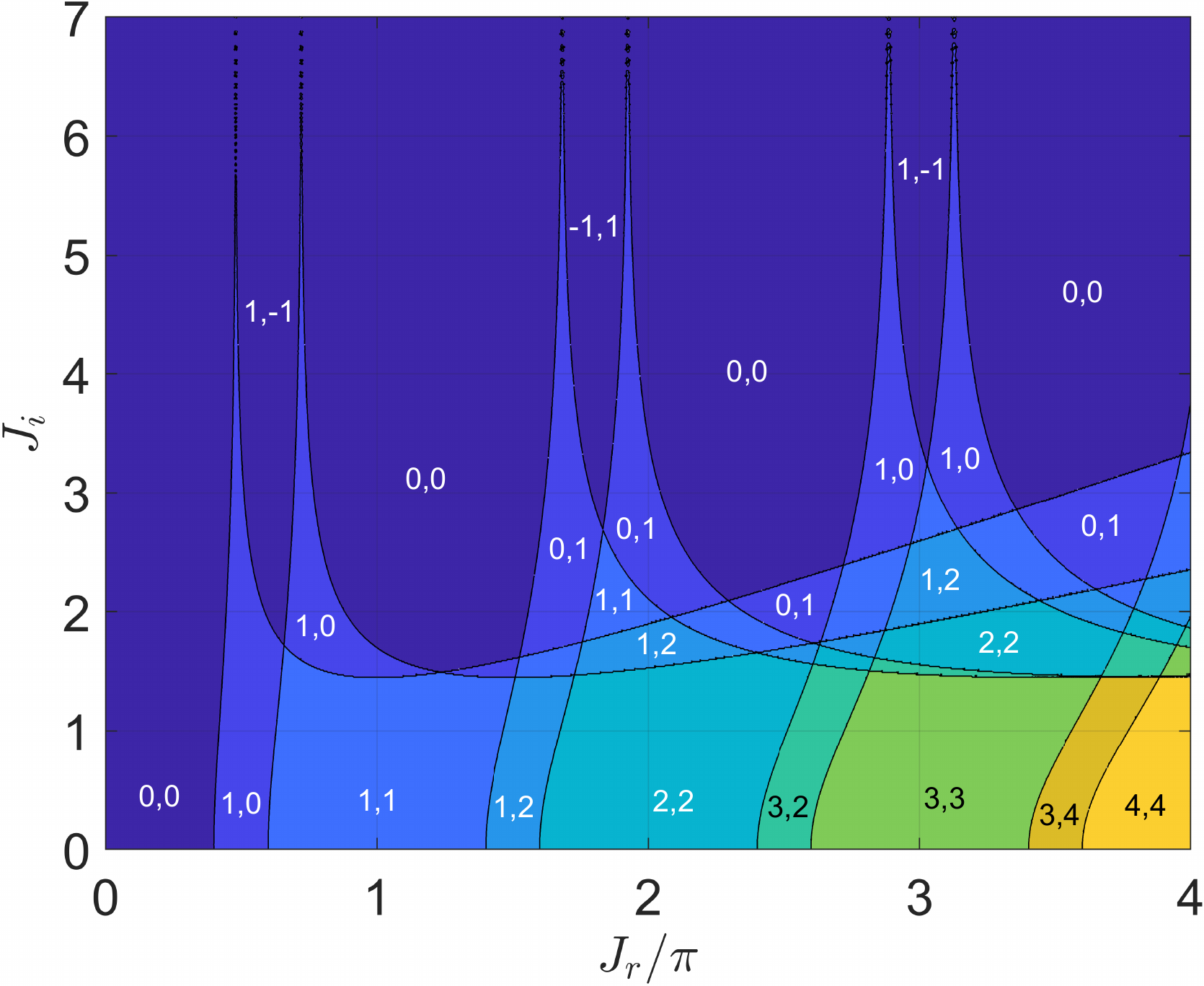}
		\par\end{centering}
	\caption{The topological phase diagram of the periodically kicked NHKC with
		respect to the real and imaginary parts of hopping amplitude $(J_{r},J_{i})$.
		The onsite chemical potential and superconducting pairing strength
		are chosen as $\mu=0.4\pi$ and $\Delta=0.9\pi$. Each region in the
		figure with a uniform color refers to a non-Hermitian Floquet superconducting
		phase, characterized by the topological winding numbers $(w_{0},w_{\pi})$
		as shown in the region. The black solid lines separating different
		regions are the phase boundaries between distinct topological phases,
		which are determined by Eq.~(\ref{eq:Gapless2}).\label{fig:PhsDigm2}}
\end{figure}
For completeness, we also present the bulk topological phase diagram
of the periodically kicked NHKC versus the real and imaginary parts
of hopping amplitude $J_{r}$ and $J_{i}$ in Fig.~\ref{fig:PhsDigm2}.
The (real) chemical potential and the superconducting pairing strength
are fixed at $\mu=0.4\pi$ and $\Delta=0.9\pi$, respectively. Similar
to Fig.~\ref{fig:PhsDigm1}, each region in Fig.~\ref{fig:PhsDigm2}
with a uniform color corresponds to a topological phase. The phase
boundaries separating different regions are denoted by black solid
lines, which are obtained analytically from Eq.~(\ref{eq:Gapless2}).
Again, we observe that with the increase of the real hopping amplitude
$J_{r}$, a series topological phase transitions happens progressively,
accompanied by gradually increased winding numbers $(w_{0},w_{\pi})$
from small to large values, as denoted within each region. Furthermore,
with the change of the imaginary hopping amplitude $J_{i}$, topological
phase transitions unique to non-Hermitian Floquet systems also happen
successively. Notably, by increasing $J_i$ from $0$ to a larger value around $J_r=0.5\pi$, we could obtain a non-Hermitian Floquet topological phase with winding numbers $(1,-1)$ from a Hermitian one with winding numbers $(1,0)$. As will be discussed in the next section, this implies that new Floquet Majorana $\pi$ edge modes could emerge under a finite amount non-Hermiticity.
These observations further exemplify the universality
and richness of non-Hermitian Floquet topological phases and phase
transitions in the periodically kicked NHKC.

In the next section, we will establish the connection between the bulk topological invariants $(w_{0},w_{\pi})$ and the Majorana zero and $\pi$ edge modes that can appear in the periodically kicked NHKC under the OBC.

\section{Majorana edge modes and bulk-edge correspondence\label{sec:Majorana}}

Majorana edge modes are one of the most important manifestations of
the topological properties of KC. They appear in the system
under the OBC when the system parameters are tuned
to the topological nontrivial regime. For the periodically kicked
KC, its Hamiltonian can be expressed under the OBC as:
\begin{alignat}{1}
H(t)= & \frac{1}{2}\sum_{n=1}^{L}\mu(2c_{n}^{\dagger}c_{n}-1)+\frac{1}{2}\sum_{n=1}^{L-1}J(c_{n}^{\dagger}c_{n+1}+{\rm H.c.})\nonumber \\
- & \frac{1}{2}\sum_{n=1}^{L-1}[\Delta\delta_{T}(t)c_{n}^{\dagger}c_{n+1}^{\dagger}+{\rm H.c.}]\label{eq:HtOBC}
\end{alignat}
where $\delta_{T}(t)\equiv\sum_{j\in\mathbb{Z}}\delta(t/T-j)$,
$L$ is the total number of lattice sites, and $H(t)$ is non-Hermitian
if either the hopping amplitude $J\in\mathbb{C}$ or the chemical
potential $\mu\in\mathbb{C}$. To find the Floquet Majorana edge modes
of the system, we need to express $H(t)$ in the Majorana representation
and obtain the corresponding equation of motion. To achieve this,
we first express fermionic annihilation and creation operators $c_{n},c_{n}^{\dagger}$
in terms of Majorana operators $\gamma_{n}^{a},\gamma_{n}^{b}$ as
\begin{equation}
c_{n}=\frac{\gamma_{n}^{a}-i\gamma_{n}^{b}}{2},\qquad c_{n}^{\dagger}=\frac{\gamma_{n}^{a}+i\gamma_{n}^{b}}{2},
\end{equation}
where $n=1,2,...,L$. The Majorana operators are Hermitian [$\gamma_{n}^{a,b}=(\gamma_{n}^{a,b})^{\dagger}$]
and satisfy the anti-commutation relation
\begin{equation}
\{\gamma_{\ell}^{s},\gamma_{m}^{s'}\}=2\delta_{\ell m}\delta_{ss'},\label{eq:MajoranaCR}
\end{equation}
where $\ell,m=1,2,...,L$ and $s,s'=a,b$. In terms of these Majorana
operators, $H(t)$ in Eq.~(\ref{eq:HtOBC}) takes the form:
\begin{alignat}{1}
H(t)= & \frac{i}{4}\sum_{n=1}^{L}(-2\mu)\gamma_{n}^{a}\gamma_{n}^{b}\nonumber \\
+ & \frac{i}{4}\sum_{n=1}^{L-1}J(\gamma_{n}^{b}\gamma_{n+1}^{a}-\gamma_{n}^{a}\gamma_{n+1}^{b})\nonumber \\
- & \frac{i}{4}\sum_{n=1}^{L-1}\Delta\delta_{T}(t)(\gamma_{n}^{b}\gamma_{n+1}^{a}+\gamma_{n}^{a}\gamma_{n+1}^{b}).\label{eq:HMt}
\end{alignat}

The Floquet operator of the system can be derived by solving the Heisenberg
equation for Majorana modes, i.e., 
\begin{equation}
\frac{d\gamma_{\ell}^{s}(t)}{dt}=i[H(t),\gamma_{\ell}^{s}(t)]\label{eq:MajoranaEOM}
\end{equation}
for $\ell=1,...,L$ and $s=a,b$, where $t$ is the scaled time under
the unit choice $\hbar=T=1$. Plugging Eq.~(\ref{eq:HMt})
into Eq.~(\ref{eq:MajoranaEOM}) and using the commutation relation (\ref{eq:MajoranaCR}),
we could express the Floquet operator $U$ of the system in Majorana
basis in the matrix form:
\begin{equation}
{\cal U}=e^{-i{\cal H}_{2}}e^{-i{\cal H}_{1}},\label{eq:UMajBas}
\end{equation}
where the explicit expressions of ${\cal H}_{1}$ and ${\cal H}_{2}$
are given in Appendix~\ref{subsec:UMaj} [alternatively, one can
express the Floquet operator in the Bogoliubov-de Gennes (BdG) representation, as elaborated
in the Appendix~\ref{subsec:UBdG}]. 
To see the symmetries that protect the Majorana edge modes of $\cal{U}$, we can first transform $\cal{U}$ to the symmetric time frames as 
\begin{alignat}{1}
{\cal U}_1 & = e^{-i{\cal H}_{1}/2}e^{-i{\cal H}_{2}}e^{-i{\cal H}_{1}/2},\label{eq:UM1}\\
{\cal U}_2 & = e^{-i{\cal H}_{2}/2}e^{-i{\cal H}_{1}}e^{-i{\cal H}_{2}/2}.\label{eq:UM2}
\end{alignat}
Then it is clear that the Floquet operators ${\cal U}_1$ and ${\cal U}_2$ in the Majorana basis possess the time-reversal symmetry ${\cal T}=\sigma_0\otimes{\mathbb I}_L$, particle-hole symmetry ${\cal C}=\sigma_z\otimes{\mathbb I}_L$ and sublattice symmetry ${\cal S}=\sigma_z\otimes{\mathbb I}_L$, i.e., ${\cal U}^{-1}_{\alpha} = {\cal T}{\cal U}^{\top}_{\alpha}{\cal T}^{-1}$, ${\cal U}_{\alpha} = {\cal C}{\cal U}^{\top}_{\alpha}{\cal C}^{-1}$, and ${\cal U}^{-1}_{\alpha}={\cal S}{\cal U}_{\alpha}{\cal S}$ for $\alpha=1,2$, where $\sigma_z$ is the Pauli matrix, $\sigma_0$ is the $2\times2$ identity matrix, ${\mathbb I}_L$ is the $L\times L$ identity matrix and ${\top}$ denotes the matrix transposition~(see Appendix \ref{subsec:UMaj}). In parallel with the Hermitian case, the Floquet Majorana modes in the periodically kicked NHKC are protected by these symmetries. 

To clearly show the Floquet Majorana edge states at both quasienergies
zero and $\pi$ in the spectrum, we introduce a pair of point-``gap functions''
$F_{0}$ and $F_{\pi}$, defined as:
\begin{alignat}{1}
F_{0}\equiv & \frac{1}{\pi}\sqrt{({\rm Re}\varepsilon)^{2}+({\rm Im}\varepsilon)^{2}},\label{eq:GapF0}\\
F_{\pi}\equiv & \frac{1}{\pi}\sqrt{(|{\rm Re}\varepsilon|-\pi)^{2}+({\rm Im}\varepsilon)^{2}},\label{eq:GapFP}
\end{alignat}
where $\varepsilon$ corresponds to all the quasienergy eigenvalues,
obtained by diagonalizing ${\cal U}$ in Eq.~(\ref{eq:UMajBas}) and
taking the logarithm of its eigenvalues. It is clear that the eigenmodes
satisfying $F_{0}=0$ ($F_{\pi}=0$) have the quasienergy $\varepsilon=0$
($\varepsilon=\pi$). If there is a point gap (on the complex plane)
between these modes and all the other eigenmodes of ${\cal U}$, we
could identify the eigenstates with $F_{0}=0$ ($F_{\pi}=0$) as Floquet
Majorana zero ($\pi$) edge modes in the periodically kicked NHKC.

\begin{figure}
	\begin{centering}
		\includegraphics[scale=0.5]{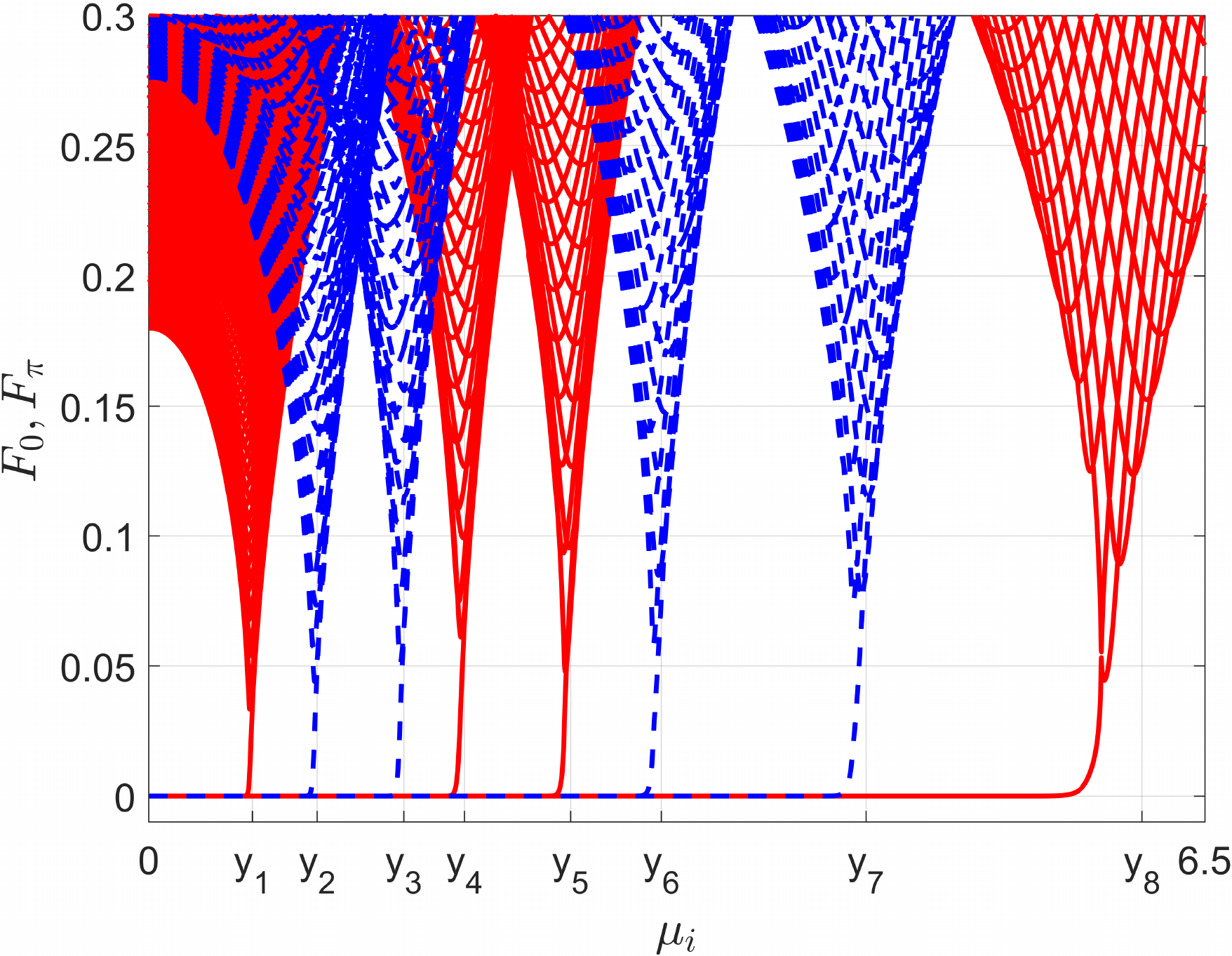}
		\par\end{centering}
	\caption{The real and imaginary parts of quasienergy spectrum [in panels (a), (b)], and 
		gap functions $F_{0}$ {[}red sold lines in panel (c), see Eq.~(\ref{eq:GapF0}){]},
		$F_{\pi}$ {[}blue dashed lines in panel (c), see Eq.~\ref{eq:GapFP}){]} of
		the periodically kicked NHKC versus the imaginary part of onsite chemical
		potential $\mu_{i}$ under OBC. The other system
		parameters are fixed at $\mu_{r}=0.3\pi$, $J=4\pi$ and $\Delta=0.5\pi$.
		The length of the lattice is $L=1000$. The solid (dashed) lines appearing
		at $F_{0}=0$ ($F_{\pi}=0$) correspond to non-Hermitian Floquet Majorana
		zero ($\pi$) modes, whose number is determined by the bulk winding
		number $w_{0}$ ($w_{\pi}$) in Eq.~(\ref{eq:W0P}). The ticks $y_{1}\sim y_{8}$
		along the $\mu_{i}$ axis in panel (c) correspond to the phase transition points
		obtained from Eq.~(\ref{eq:Gapless1}). Their numerical values are
		approximately given by $\mu_{i}\approx(0.64,1.03,1.57,1.94,2.60,3.15,4.41,6.11)$,
		respectively.\label{fig:SpectrumOBC1}}
\end{figure}
In Fig.~\ref{fig:SpectrumOBC1}, we show the quasienergy spectrum and gap functions $F_{0}$
(red solid lines), $F_{\pi}$ (blue dashed lines) of the periodically
kicked NHKC versus the imaginary part of the chemical potential $\mu_{i}$
under OBC. The real part of chemical potential,
hopping amplitude, superconducting pairing strength and the number
of lattice sites are fixed at $\mu_{r}=0.3\pi$, $J=4\pi$, $\Delta=0.5\pi$
and $L=1000$, respectively. The values of $\mu_{i}$ at grids $y_{1}\sim y_{8}$
are computed analytically from the gapless condition Eq.~(\ref{eq:Gapless1}).
They correspond to the bulk topological phase transition points of
the periodically kicked NHKC. We notice that across each transition
point, a pair of Floquet Majorana zero modes (red solid lines) or $\pi$
modes (blue dashed lines) merge into the bulk. Referring to the phase
diagram Fig.~\ref{fig:PhsDigm1}, we further observe that each of the above mentioned
transitions is accompanied by a quantized change of winding number
$w_{0}$ or $w_{\pi}$ by $1$. Between each pair of adjacent topological
transitions, the winding numbers $(w_{0},w_{\pi})$ also predict the
number of degenerate Majorana edge modes $(N_{0},N_{\pi})$ at quasienergies
zero and $\pi$, as $N_{0}=2|w_{0}|$ and $N_{\pi}=2|w_{\pi}|$. These
relations establish the correspondence between the bulk topological
invariants and the Floquet topological edge states in the periodically
kicked NHKC.
From the perspective of symmetry classification, we notice that the effective Hamiltonian of our system in each time frame $\alpha$ possesses the time-reversal symmetry in Eq.~(\ref{eq:HaTRST}) and parity symmetry as $\sigma_x H_\alpha(k)\sigma_x=H_\alpha(-k)$. According to the analysis in Ref.~\cite{Class7}, the non-Hermitian skin effect can be avoided and the conventional type of bulk-edge correspondence can be recovered when the Hamiltonian of a static non-Hermitian system has these symmetries. Our results demonstrate that the analysis in Ref.~\cite{Class7} about the recovery of bulk-edge correspondence can also be carried over to Floquet systems at the level of effective Hamiltonians.
Note in passing that the deviations between theory and numerical results around $\mu_{i}=y_{7},y_{8}$ are finite-size effects, which will tend to vanish in the thermodynamic limit $L\rightarrow\infty$~(see Appendix~\ref{subsec:FSE} for a demonstration).

\begin{figure}
	\begin{centering}
		\includegraphics[scale=0.5]{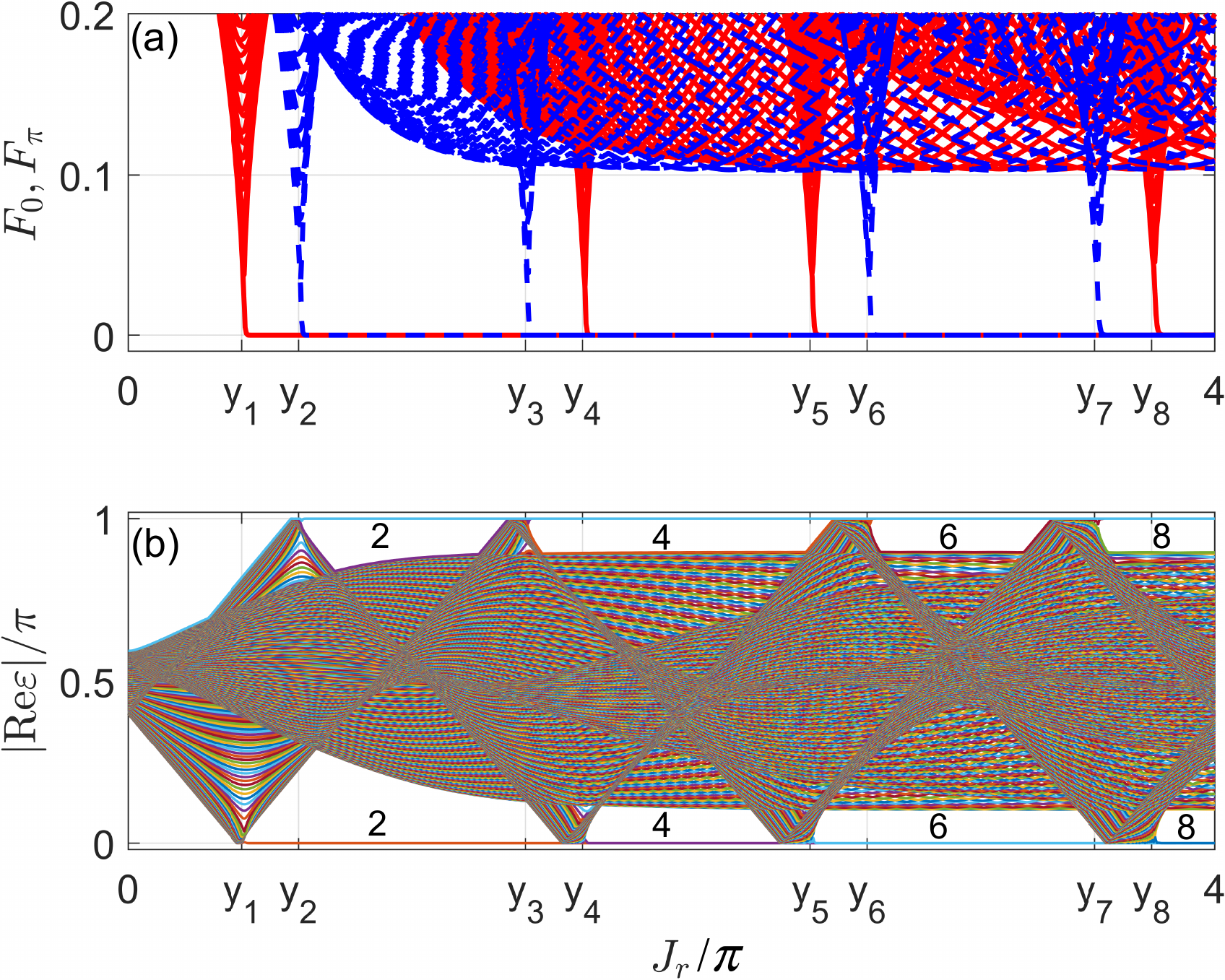}
		\par\end{centering}
	\caption{The gap functions $F_{0}$ {[}red sold lines, see Eq.~(\ref{eq:GapF0}){]},
		$F_{\pi}$ {[}blue dashed lines, see Eq.~\ref{eq:GapFP}){]}, and
		the real part of quasienergy $\varepsilon$ of the periodically kicked
		NHKC versus the real part of hopping amplitude $J_{r}$ under OBC. The other system parameters are fixed at $\mu=0.4\pi$,
		$\Delta=0.9\pi$, $J_{i}=1$, and $L=500$ for both panels (a) and
		(b). The solid (dashed) lines appearing at $F_{0}=0$ ($F_{\pi}=0$)
		in panel (a) correspond to non-Hermitian Floquet Majorana zero ($\pi$)
		modes, whose number is determined by the bulk winding number $w_{0}$
		($w_{\pi}$) in Eq.~(\ref{eq:W0P}). The ticks $y_{1}\sim y_{8}$
		along the $J_{r}$ axis in both panels correspond to the phase transition
		points obtained from Eq.~(\ref{eq:Gapless1}). Their numerical values
		are approximately given by $J_{r}/\pi\approx(0.42,0.63,1.47,1.68,2.51,2.72,3.56,3.77)$,
		respectively. The numerical values in panel (b) denote the number
		of Floquet Majorana edge modes at quasienergies zero and $\pi$ in
		each gapped regions.\label{fig:SpectrumOBC2}}
\end{figure}
\begin{figure}
	\begin{centering}
		\includegraphics[scale=0.48]{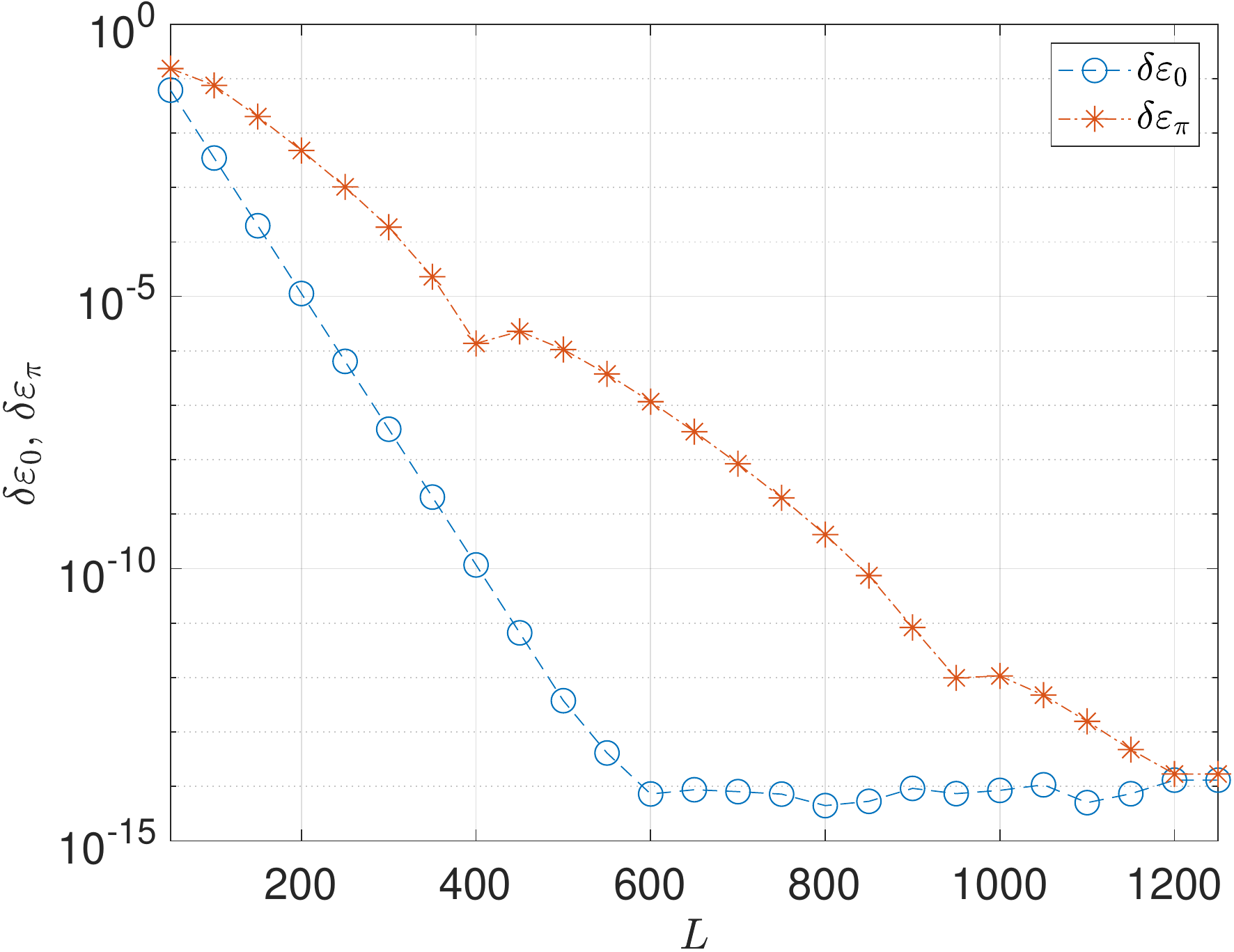}
		\par\end{centering}
	\caption{Scaling of the largest quasienergy splittings among non-Hermitian
		Floquet Majorana zero modes ($\delta\varepsilon_{0}$, blue circles)
		and $\pi$ modes ($\delta\varepsilon_{\pi}$, red stars) versus the
		lattice size $L$ of the periodically kicked NHKC. The system parameters
		are chosen to be $J=4\pi$, $\mu=0.3\pi+1.3i$, and $\Delta=0.5\pi$.
		Note that beyond $L=600$, the maximal level spacing among Majorana
		zero modes already reaches the machine accuracy.\label{fig:MajoranaScaling}}
\end{figure}

For completeness, in Fig.~\ref{fig:SpectrumOBC2} we also present
the real part of Floquet quasienergy spectrum and the gap functions of the periodically kicked
NHKC with non-Hermitian hopping amplitudes ($J_{i}\neq0$), and observed
similar results as in the case of onsite gain/loss ($\mu_{i}\neq0$).
Note in passing that in Fig.~\ref{fig:SpectrumOBC2}(b), multiple
pairs of degenerate Majorana zero and $\pi$ modes could appear and
coexist in the same parameter regime, as denoted in the figure. These
Majorana modes could have applications in the recently proposed concept
of Floquet topological quantum computing~\cite{FloCompt1,FloCompt2,FloCompt3}, with a potential advantage
due to their persistence in the non-Hermitian regions.
\begin{figure}
	\begin{centering}
		\includegraphics[scale=0.48]{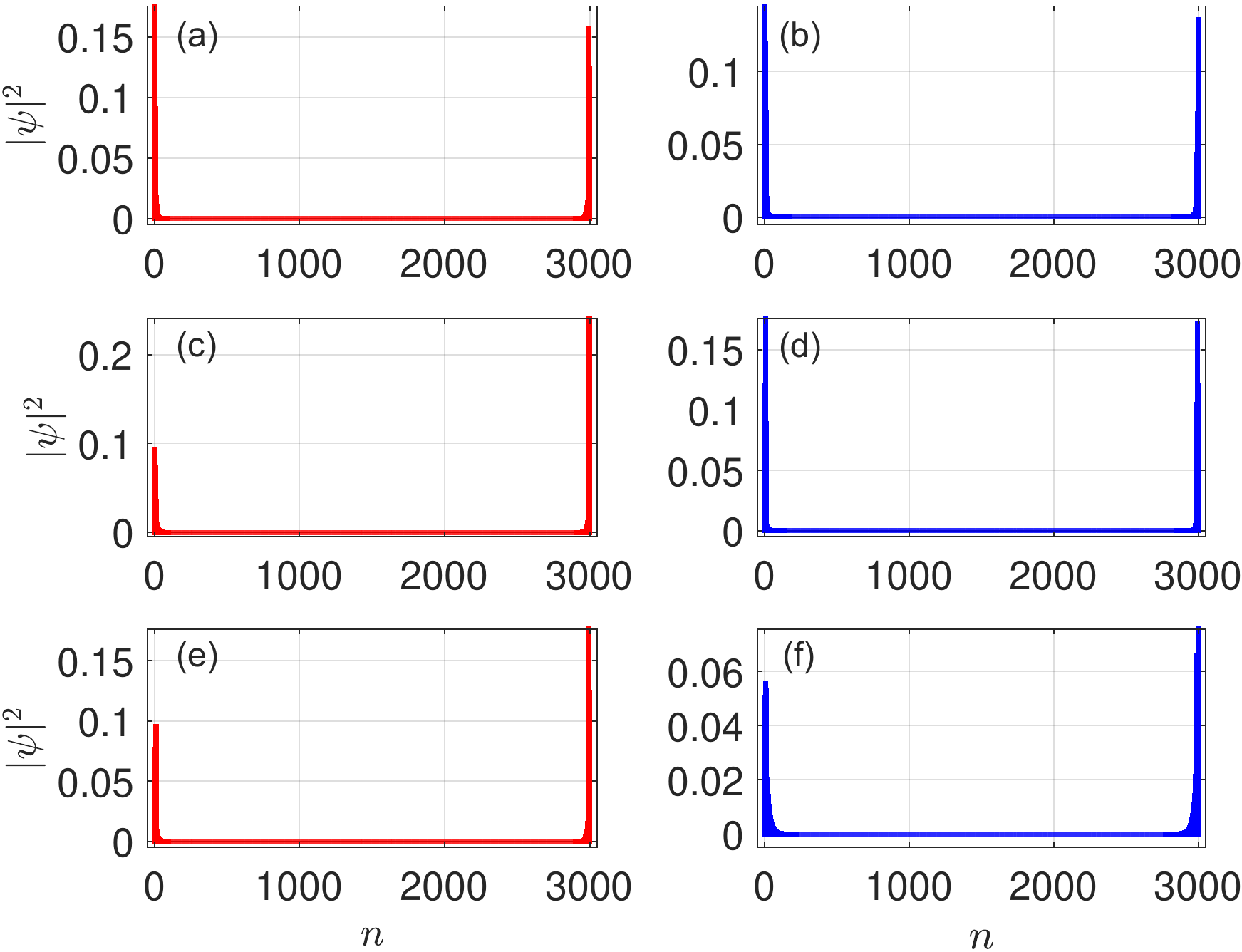}
		\par\end{centering}
	\caption{Probability distributions $|\psi|^{2}$ of the Majorana zero modes
		{[}red columns in panels (a), (c), (e){]} and $\pi$ modes {[}blue
		columns in panels (b), (d), (f){]} versus the lattice site index $n$
		in the periodically kicked NHKC. The system parameters are chosen
		to be $J=4\pi$, $\mu=0.3\pi+1.3i$, $\Delta=0.5\pi$, and the length
		of the chain $L=3000$. There are six Majorana edge
		modes at both quasienergies zero and $\pi$, as predicted by the bulk
		winding numbers $(w_{0},w_{\pi})$ in Eq.~(\ref{eq:W0P}).\label{fig:MajoranaEM}}
\end{figure}
From Figs.~\ref{fig:SpectrumOBC1} and \ref{fig:SpectrumOBC2}, we
see that changing the gain and loss in the onsite chemical potential or the hopping
amplitude can indeed induce topological phase transitions in the periodically
kicked NHKC, and create non-Hermitian topological superconducting
phases that are unique to Floquet systems. These phases are further characterized
by the appearance of non-Hermitian Floquet Majorana zero and $\pi$
edge modes under the OBC. One example of the spatial profiles of
these Majorana modes is shown explicitly in Fig.~\ref{fig:MajoranaEM},
where we see that the Majorana zero modes [red columns in Figs.~\ref{fig:MajoranaEM}(a,c,e)] and $\pi$ modes [blue columns in Figs.~\ref{fig:MajoranaEM}(b,d,f)] are indeed localized exponentially around the left and right edges of the chain. In Fig.~\ref{fig:MajoranaScaling},
we further show the largest quasienergy splittings of the Majorana zero and $\pi$
modes versus the length $L$ of the chain. We observe that both the maximal
splittings $\delta\varepsilon_{0}$ and $\delta\varepsilon_{\pi}$
of the Majorana zero and $\pi$ modes decrease exponentially with the
increase of the system size $L$ before reaching the machine precision, which is expected for Majorana edge modes.

Putting together, we conclude that the periodically kicked NHKC could indeed possess multiple Floquet Majorana zero and $\pi$ edge modes under OBC. The number of these modes in each topological phase is correctly counted by the bulk winding numbers we introduced in Eq.~(\ref{eq:W0P}).

Before discussing the experimental relevance of our results, we would like to highlight a couple of aspects that distinguish our work from previous studies on non-Hermitian Floquet matter~\cite{ZhouPRB2019,ZhouPRB2018,ZhouPRA2019,FehskePRL2019,LiPRB2019,YuceEPJD2015,NHSkin12,NHPump1}. The studies in \cite{ZhouPRB2019,ZhouPRB2018,ZhouPRA2019} focus on the engineering of non-Hermitian Floquet topological insulators in 1d and the dynamical detection of their topological invariants. The systems considered there belong to the symmetry class AIII or BDI, which are also characterized by a pair of integer winding numbers. However, superconducting systems with non-Hermitian Floquet Majorana modes have not been considered in \cite{ZhouPRB2019,ZhouPRB2018,ZhouPRA2019}. In another set of studies \cite{LiPRB2019,FehskePRL2019,NHSkin12,NHPump1}, non-Hermitian Floquet phases in two-dimensional systems have also been considered, with a focus on the chiral edge states, non-Hermitian skin effects and their related transport phenomena. However, the model studied in the current work has a couple of symmetries and therefore is immune to the non-Hermitian skin effect, which distinguishes it from the situations considered in \cite{LiPRB2019,FehskePRL2019,NHSkin12,NHPump1}. Besides these differences, and to the best of our knowledge, our work provides the first demonstration of a non-Hermitian Floquet topological superconducting model with multiple Majorana edge modes, and confirmed the bulk-edge correspondence therein. Furthermore, we have extended a couple of techniques (e.g., the description of time-periodic dynamics in Majorana and BdG basis) to non-Hermitian Floquet systems, which are applied previously in Hermitian settings. These extensions will benefit further theoretical and numerical explorations of non-Hermitian Floquet superconducting phases in different symmetry classes and across different physical dimensions.

\vspace{0.5cm}

\section{Connection to experiments\label{sec:Exp}}

In this section, we discuss a couple of issue regarding the experimental realization of our periodically kicked NHKC model and the detection of its topological phases.

One of the candidate platforms in which our model may be realized is the cold atom system, in which the engineering of Hermitian Floquet KC has been discussed~\cite{FloMaj0,FloMaj8}. In our non-Hermitian model, the imaginary part of chemical potential $\mu_i$ corresponds to onsite atom loss in cold atom systems, and its magnitude corresponds to the loss rate. To realize and control such a loss, one may consider using a resonant optical beam to kick the atoms out of a trap, which has been experimentally realized in $^6$Li~\cite{NHKCExp3}. Alternatively, one may consider applying a radio-frequency~(rf) pulse to excite atoms to another irrelevant state $|i\rangle$, leading to an effective decay when atoms in $|i\rangle$ experience a loss by applying an anti-trap~\cite{AntiTrap}. Similarly, the imaginary part of hopping amplitude $J_i$ corresponds to the loss of atoms when they hop between neighboring sites, and its magnitude controls the corresponding loss rate. In experiments, $J_i$ may also be introduced by a resonant optical beam or a rf pulse, which is now applied to the atoms when they tunnel between neighboring sites. Another possible way of engineering $J_i$ is to apply and extend the techniques introduced in the realization of asymmetric hopping for cold atoms in Ref.~\cite{NHTPReview1}.

In cold atom systems, the $p$-wave superconducting pairing term can be introduced by combining orbital degrees of freedom with strong s-wave interactions~\cite{PairingExp1}, via the synthetic spin-orbit coupling~\cite{SynSOC}, or through the $p$-wave Feshbach resonance~\cite{FeshbachRes}. To realize a periodically kicked superconducting pairing term, one may control the Feshbach resonance by a time-dependent magnetic field~\cite{FeshbachRes2} or laser field~\cite{FeshbachRes3}, and tailor the magnetic or laser field used for the tuning to a short-pulsed modulation with a large amplitude, which may then be effectively modeled by $\delta$-kicks. If the pairing term is realized by synthetic spin-orbit coupling, it may also be controlled by a periodic modulation of the Raman-coupling amplitude~(i.e., tuning the corresponding laser intensity periodically in time)~\cite{SynSOC2}. By shaping each modulation to a short-pulsed form with a large amplitude, the pairing term originated from this type of synthetic spin-orbit coupling may also be modeled by $\delta$-kicks.

In cold atom systems, the winding numbers of our periodically kicked NHKC may be detected by measuring its dynamical pseudo-spin textures in momentum space, from which a pair of dynamical winding numbers may be constructed and shown to be equal to the topological invariants of the system. Such a dynamical scheme has been proposed already in both non-Hermitian static~\cite{DynamicWN} and Floquet~\cite{ZhouPRB2019} systems with chiral symmetry. Another possible way of detecting the topological phases in our model is by measuring the mean chiral displacement (MCD) of a wave packet prepared initially at the center of the lattice. In a non-Hermitian Floquet system with chiral symmetry in 1d, it has also been shown that the long-time averaged MCD could produce the topological winding numbers of the system~\cite{ZhouPRA2019}. In recent years, the observation of both dynamical spin textures~\cite{DynamicST} and MCDs~\cite{MCDExp} have been achieved in cold atom systems.

Finally, to probe the non-Hermitian Floquet Majorana edge modes in our system, one may consider to apply the spatially resolved rf spectroscopy. A possible idea is to induce a single particle excitation from a fermionic state to an unoccupied fluorescent probe state by a probe rf field. Spatial information about the local density of states can then be gained by imaging the population in the probe state. Ionization or in situ imaging techniques may further be used to readout the population in the probe state with single particle resolution~\cite{FloMaj0}. However, as there are multiple pairs of zero and $\pi$ Floquet Majorana edge modes in our system, the final confirmation of their exact numbers in experiments would be a challenging task.

Note in passing that the system parameters $(\mu,J,\Delta)$ used in our presentations of Figs.~\ref{fig:PhsDigm1} to \ref{fig:SpectrumOBC2} are all in dimensionless units, and the unit of energy is chosen as $\hbar/T$ or $\hbar\omega/2\pi$, with $T=2\pi/\omega$ and $\omega$ being the driving frequency. Compared with the energy scale of driving photons (i.e., $\hbar\omega$), our system parameters are taken mainly in the range $0.1\hbar\omega\sim2\hbar\omega$ in all the Figures. Since a cold atom system usually has a relatively long coherent time, one would have more freedom to tune the driving frequency $\omega$ in experiments. Therefore, an appropriate choice of $\omega$ should allow all other system parameters to fall into the regime of our interest. In earlier studies of Floquet topological superconductors~\cite{FloMaj0,FloMaj8}, it was mentioned that an experimentally suitable range of driving frequency would be around $\omega\in(62,150)(2\pi)$kHz, and the other systems parameters can be set to either below or nearby this range. Therefore, we believe that most of the system parameters we used to obtain the phase diagrams and the spectrum of our model should be within reach in current cold atom experiments.

\vspace{0.5cm}

\section{Summary and discussion\label{sec:Summary}}

In this work, we found rich non-Hermitian Floquet topological superconducting
phases in a periodically kicked non-Hermitian Kitaev chain. Our system
belongs to an extended BDI class in non-Hermitian Floquet systems. Each of
its topological phases is characterized by a pair of integer winding
numbers, which can take large values due to the interplay between
driving and non-Hermitian effects. Under OBC,
multiple Floquet Majorana modes with quasienergies zero and $\pi$
appear at the edges of the chain, with their numbers being determined by
the topological invariants of the bulk Floquet operator. The degeneracy
of these edge modes is protected by the symmetries that
are unique to non-Hermitian systems. These results establish a new
class of Floquet topological superconductors in non-Hermitian settings.

A potential application of our discovery resides in the recently proposed
Floquet topological quantum computing~\cite{FloCompt1,FloCompt2,FloCompt3}. There, time is utilized as
an extra dimension to assist the braiding of Majorana zero and $\pi$
modes at the edges or corners of Floquet topological superconductors. The non-Hermitian
Floquet Majorana modes found in this work might be able to make the Floquet topological
quantum computations more robust to environmental-induced nonreciprocity,
losses, and certain types of quasiparticle poisoning effects. On the other hand, the existence of more than one pairs of non-Hermitian
Floquet Majorana modes could create stronger signals at the edges
of the system, making it easier for the experimental detection of
their topological signatures in open system settings.


\section*{Acknowledgement}
L.Z. acknowledges R.W. Bomantara and J. Gong for helpful comments. This work is supported by the National Natural Science Foundation of China (Grant No.~11905211), the China Postdoctoral Science Foundation (Grant No.~2019M662444), the Fundamental Research Funds for the Central Universities (Grant No.~841912009), the Young Talents Project at Ocean University of China (Grant No.~861801013196), and the Applied Research Project of Postdoctoral Fellows in Qingdao (Grant No.~861905040009).

\appendix
\section{Derivation of Eq.~(\ref{eq:UkPKNHKC}) from Eq.~(\ref{eq:Hkt})\label{subsec:Uk}}
The Floquet operator of our model is given by its time-evolution operator over a complete driving period (e.g. from $t=nT+0^-$ to $t=nT+T+0^-$ with $n\in{\mathbb Z}$), i.e., $U(k)={\mathbb T}e^{-i\int_{nT+0^-}^{(n+1)T+0^-}H(k,t)dt}$, where ${\mathbb T}$ performs the time ordering. The time-dependent Hamiltonian $H(k,t)$ is given by Eq.~(\ref{eq:Hkt}), in which the function $\delta_T(t)$ is defined as $\delta_{T}(t)\equiv\sum_{\ell\in\mathbb{Z}}\delta(t/T-\ell)$. Plugging $H(k,t)$ into the integral, we see that in the time duration $(nT+0^-,nT+0^+)$, the evolution is determined by $U_y(k)=e^{\frac{-i}{\hbar}\int_{nT+0^-}^{nT+0^+}h_{y}(k)\delta_{T}(t)\sigma_{y}dt}=e^{\frac{-i}{\hbar}\int_{nT+0^-}^{nT+0^+}h_{y}(k)\delta(t/T-n)\sigma_{y}dt}=e^{\frac{-i}{\hbar}T\int_{n+0^-}^{n+0^+}h_{y}(k)\delta(s-n)\sigma_{y}ds}$, where to arrive at the last equality we have introduced the scaled time $s=t/T$. Working out the integral then yields $U_y(k)=e^{-i\frac{T}{\hbar}h_{y}(k)\sigma_{y}}$. Similarly, in the time duration $(nT+0^+,nT+T+0^-)$, the dynamics of the system is governed by the Hamiltonian $h_z(k)\sigma_z$, and the corresponding evolution operator is $U_z(k)=e^{-i\frac{T}{\hbar}h_z(k)\sigma_z}$. Putting together and setting $\hbar=T=1$, we find the Floquet operator of the system in momentum space to be $U(k)=U_z(k)U_y(k)$, as given by Eq.~(\ref{eq:UkPKNHKC}) in the main text.

\vspace{0.5cm}

\section{Analytical derivation of the gapless conditions\label{subsec:GaplessCond}}

In this Appendix, we give the derivation details for the gapless conditions
of the periodically kicked NHKC. According to the main text, the Floquet
spectrum of the periodically kicked NHKC becomes gapless at the center
or boundary of the quasienergy Brillouin zone if the dispersion relation
$\varepsilon(k)=0$ or $\pi$, respectively. Putting together,
we find that the bulk's spectrum gap closes if
\begin{equation}
\cos[h_{y}(k)]\cos[h_{z}(k)]=\pm1,\label{eq:GapCondApp}
\end{equation}
where $h_{y}(k)=\Delta\sin k$ and $h_{z}(k)=\mu+J\cos k$ as defined
in the main text. When both the chemical potential $\mu$ and hopping
amplitude $J$ take real values, it is straightforward to see that
the condition (\ref{eq:GapCondApp}) is met if
\begin{equation}
\left(\frac{m\pi}{\Delta}\right)^{2}+\left(\frac{n\pi-\mu}{J}\right)^{2}=1,\label{eq:GapCondReal}
\end{equation}
where $m,n\in\mathbb{Z}$, $|m\pi|\leq|\Delta|$ and $|n\pi-\mu|\leq|J|$.
Eq.~(\ref{eq:GapCondReal}) determines the gapless conditions of the
periodically kicked Hermitian KC in the space of system
parameters $\Delta$, $\mu$ and $J$. When either $\mu\in\mathbb{C}$
or $J\in\mathbb{C}$, the gapless conditions are different from Eq.~(\ref{eq:GapCondReal}). We discuss the cases with $(\mu\in\mathbb{C},J\in\mathbb{R})$
and $(J\in\mathbb{C},\mu\in\mathbb{R})$ below separately.

\subsection{$\mu\in\mathbb{C}$ and $J\in\mathbb{R}$}

In this case, the system is subject to onsite gain/losses. We denote the real and imaginary parts of the chemical potential
$\mu$ as $\mu_{r}$ and $\mu_{i}$, such that $\mu=\mu_{r}+i\mu_{i}$.
The gapless condition (\ref{eq:GapCondApp}) then becomes
\begin{alignat}{1}
\cos(\Delta\sin k)\cos(\mu_{r}+J\cos k)\cosh\mu_{i}= & \pm1\label{eq:GapCond11},\\
\cos(\Delta\sin k)\sin(\mu_{r}+J\cos k)\sinh\mu_{i}= & 0\label{eq:GapCond12}.
\end{alignat}
As $\mu_{i}\neq0$, we have $\sinh\mu_{i}\neq0$ in Eq.~(\ref{eq:GapCond12}).
Moreover, the condition Eq.~(\ref{eq:GapCond11}) cannot be satisfied
when $\cos(\Delta\sin k)=0$. Therefore, to satisfy both Eqs.~(\ref{eq:GapCond11})
and (\ref{eq:GapCond12}), we must have $\sin(\mu_{r}+J\cos k)=0$,
yielding $\mu_{r}+J\cos k=n\pi$ for $n\in\mathbb{Z}$. The solution
of this equation determines the quasimomentum $k$ at which the band
touches, i.e.,
\begin{equation}
\cos k=\frac{n\pi-\mu_{r}}{J}\qquad{\rm for}\qquad|n\pi-\mu_{r}|\leq|J|.\label{eq:cosk1}
\end{equation}
In the meantime, $\sin(\mu_{r}+J\cos k)=0$ also implies $\cos(\mu_{r}+J\cos k)=\pm1$.
Plugging this and Eq.~(\ref{eq:cosk1}) into Eq.~(\ref{eq:GapCond11})
finally yields
\begin{equation}
\mu_{i}=\pm{\rm arccosh}\left\{ \frac{1}{\left|\cos\left[\Delta\sqrt{1-(n\pi-\mu_{r})^{2}/J^{2}}\right]\right|}\right\} .\label{eq:GaplessApp1}
\end{equation}
Therefore, under the condition $n\in\mathbb{Z}$ and $|n\pi-\mu_{r}|\leq|J|$,
the Floquet spectrum of periodically kicked NHKC is gapless when the Eq.~(\ref{eq:GaplessApp1}) for the system parameters is satisfied, as also given by Eq.~(\ref{eq:Gapless1}) in the main text.

\subsection{$J\in\mathbb{C}$ and $\mu\in\mathbb{R}$}

In this case, the system is subject to non-Hermitian hoppings.
We denote the real and imaginary parts of hopping amplitude
$J$ as $J_{r}$ and $J_{i}$, such that $J=J_{r}+iJ_{i}$. The gapless
condition (\ref{eq:GapCondApp}) then becomes
\begin{alignat}{1}
\cos(\Delta\sin k)\cos(\mu+J_{r}\cos k)\cosh(J_{i}\cos k)= & \pm1\label{eq:GapCond21},\\
\cos(\Delta\sin k)\sin(\mu+J_{r}\cos k)\sinh(J_{i}\cos k)= & 0\label{eq:GapCond22}.
\end{alignat}
Since $J_{i}\neq0$, we must have $k=\pm\pi/2$ if $J_{i}\cos k=0$.
According to Eq.~(\ref{eq:GapCond21}), this further implies that
$\Delta=m\pi$ and $\mu=n\pi$ for $m,n\in\mathbb{Z}$. These conditions
only give isolated points in the parameter space, instead of continuous
boundary lines between possibly different phases. Also, the imaginary
part of hopping amplitude $J_{i}$ becomes irrelevant to these gapless
points, which is not the situation of our interest. When $J_{i}\cos k\neq0$,
we need to have $\sin(\mu+J_{r}\cos k)=0$ for both conditions (\ref{eq:GapCond21})
and (\ref{eq:GapCond22}) to be satisfied, yielding $\mu+J_{r}\cos k=n\pi$
for $n\in\mathbb{Z}$. The solution of this equation determines the
quasimomentum $k$ at which the band touches, i.e.,
\begin{equation}
\cos k=\frac{n\pi-\mu}{J_{r}}\qquad{\rm for}\qquad|n\pi-\mu|\leq|J_{r}|.\label{eq:cosk2}
\end{equation}
Furthermore, $\sin(\mu+J_{r}\cos k)=0$ also implies $\cos(\mu+J_{r}\cos k)=\pm1$.
Combining this and Eq.~(\ref{eq:cosk2}) into Eq.~(\ref{eq:GapCond21})
finally leads to
\begin{equation}
J_{i}=\pm\frac{J_{r}}{n\pi-\mu}{\rm arccosh}\left\{ \frac{1}{\left|\cos\left[\Delta\sqrt{1-(n\pi-\mu)^{2}/J_{r}^{2}}\right]\right|}\right\} .\label{eq:GaplessApp2}
\end{equation}
Therefore, under the condition $n\in\mathbb{Z}$ and $|n\pi-\mu|\leq|J_{r}|$,
the Floquet spectrum of periodically kicked NHKC is gapless when Eq.~(\ref{eq:GaplessApp2}) for the system parameters is satisfied, as also given by Eq.~(\ref{eq:Gapless2}) in the main text.

\section{Prove of the realness of $w_\alpha$ in Eq.~(\ref{eq:W12})}\label{subsec:ReW12}
In this appendix, we show that the imaginary part of the integrand in Eq.~(\ref{eq:W12}) will vanish after the integration, and $w_\alpha$ is indeed a well-defined winding number.

To do so, we first introduce an azimuthal angle as $\phi_\alpha(k)\equiv \arctan[h_{\alpha y}(k)/h_{\alpha z}(k)]$, such that
\begin{equation}
\partial_k\phi_\alpha(k)=\frac{h_{\alpha z}(k)\partial_k h_{\alpha y}(k)-h_{\alpha y}(k)\partial_k h_{\alpha z}(k)}{h^2_{\alpha y}(k)+h^2_{\alpha z}(k)}\label{eq:dkPhiAk}
\end{equation}
Therefore, according to Eq.~(\ref{eq:W12}), the winding number $w_\alpha$ can be expressed equivalently through $\phi_\alpha(k)$ as
\begin{alignat}{1}
w_\alpha= & \int_{-\pi}^{\pi}\frac{dk}{2\pi}\partial_k\phi_\alpha(k)\nonumber\\
= & \int_{-\pi}^{\pi}\frac{dk}{2\pi}\{\partial_k{\rm Re}[\phi_\alpha(k)]+i\partial_k{\rm Im}[\phi_\alpha(k)]\}\label{eq:WAinPHIA}.
\end{alignat}
On the other hand, we have
\begin{alignat}{1}
e^{i2\phi_{\alpha}(k)}& = e^{i2{\rm Re}[\phi_{\alpha}(k)]}e^{-2{\rm Im}[\phi_{\alpha}(k)]}\nonumber \\
& = \frac{e^{i\phi_{\alpha}(k)}}{e^{-i\phi_{\alpha}(k)}}=\frac{\cos[\phi_{\alpha}(k)]+i\sin[\phi_{\alpha}(k)]}{\cos[\phi_{\alpha}(k)]-i\sin[\phi_{\alpha}(k)]}\nonumber \\
& = \frac{1+i\tan[\phi_{\alpha}(k)]}{1-i\tan[\phi_{\alpha}(k)]}=\frac{h_{\alpha z}(k)+ih_{\alpha y}(k)}{h_{\alpha z}(k)-ih_{\alpha y}(k)}.\label{eq:ei2PPhiA}
\end{alignat}
Taking the absolute values on both sides of Eq.~(\ref{eq:ei2PPhiA}), we find that the imaginary part of azimuthal angle $\phi_\alpha(k)$ satisfy $e^{-2{\rm Im}[\phi_{\alpha}(k)]}=|h_{\alpha z}(k)+ih_{\alpha y}(k)|/|h_{\alpha z}(k)-ih_{\alpha y}(k)|$, which implies that
\begin{equation}
{\rm Im}[\phi_{\alpha}(k)]=-\frac{1}{2}\ln\left|\frac{h_{\alpha z}(k)+ih_{\alpha y}(k)}{h_{\alpha z}(k)-ih_{\alpha y}(k)}\right|.\label{eq:ImPhiA}
\end{equation}
Since $h_{\alpha z}(k)$ and $h_{\alpha y}(k)$ are both continuous and periodic functions of $k$, ${\rm Im}[\phi_{\alpha}(k)]$ in Eq.~(\ref{eq:ImPhiA}) is a real, continuous, and periodic function of $k$ with the period $2\pi$. Therefore, the integral over the imaginary part of $\partial_k\phi_\alpha(k)$ in Eq.~(\ref{eq:WAinPHIA}) yields
\begin{equation}
i\int_{-\pi}^{\pi}\frac{dk}{2\pi}\partial_{k}{\rm Im}[\phi_{\alpha}(k)]=\frac{1}{4\pi i}\left.\ln\left|\frac{h_{\alpha z}(k)+ih_{\alpha y}(k)}{h_{\alpha z}(k)-ih_{\alpha y}(k)}\right|\right|_{-\pi}^{\pi}=0,
\end{equation}
and we conclude that the imaginary part of the integrand in the definition of $w_\alpha$ in Eq.~(\ref{eq:W12}) indeed has no contribution to its final value after integrating over the whole Brillouin zone. This then confirms our statement that the imaginary part Im$[h_{\alpha y}(k),h_{\alpha z}(k)]$ of the complex vector $[h_{\alpha y}(k),h_{\alpha z}(k)]$ has no winding around its own zero. For static (non-Floquet) systems, a similar proof of the realness of winding numbers can also be found in Ref.~\cite{DynamicWN}.

\section{Floquet operator in Majorana representation\label{subsec:UMaj}}

In this appendix, we present the derivation details for the Floquet
operator of the periodically kicked NHKC in the Majorana representation.
We first write the Hamiltonian $H(t)$ in Eq.~(\ref{eq:HMt}) as $H(t)=\frac{i}{4}{\cal H}(t)$,
where 
\begin{alignat}{1}
{\cal H}(t)\equiv & (-2\mu)\sum_{n}\gamma_{n}^{a}\gamma_{n}^{b}\nonumber \\
+ & J\sum_{n}(\gamma_{n}^{b}\gamma_{n+1}^{a}-\gamma_{n}^{a}\gamma_{n+1}^{b})\nonumber \\
- & \Delta\delta_{T}(t)\sum_{n}(\gamma_{n}^{b}\gamma_{n+1}^{a}+\gamma_{n}^{a}\gamma_{n+1}^{b}).\label{eq:calHMt}
\end{alignat}
Using the commutation relation Eq.~(\ref{eq:MajoranaCR}) and the commutator formula
\begin{equation}
[AB,C]=A\{B,C\}-\{A,C\}B,\label{eq:ComABC}
\end{equation}
it is straightforward to show that
\begin{alignat}{1}
[\gamma_{n}^{a}\gamma_{n}^{b},\gamma_{\ell}^{s}]= & 2(\delta_{\ell,n}\delta_{s,b}\gamma_{\ell}^{a}-\delta_{\ell,n}\delta_{s,a}\gamma_{\ell}^{b}),\\{}
[\gamma_{n}^{b}\gamma_{n+1}^{a},\gamma_{\ell}^{s}]= & 2(\delta_{\ell,n+1}\delta_{s,a}\gamma_{\ell-1}^{b}-\delta_{\ell,n}\delta_{s,b}\gamma_{\ell+1}^{a}),\\{}
[\gamma_{n}^{a}\gamma_{n+1}^{b},\gamma_{\ell}^{s}]= & 2(\delta_{\ell,n+1}\delta_{s,b}\gamma_{\ell-1}^{a}-\delta_{\ell,n}\delta_{s,a}\gamma_{\ell+1}^{b}).
\end{alignat}
Combining these and Eq.~(\ref{eq:calHMt}) with Eq.~(\ref{eq:MajoranaEOM}),
we obtain
\begin{alignat*}{1}
\frac{d\gamma_{\ell}^{s}(t)}{dt}= & \mu\delta_{s,b}\gamma_{\ell}^{a}(t)-\mu\delta_{s,a}\gamma_{\ell}^{b}(t)\\
- & \frac{J}{2}\delta_{s,a}\left[\gamma_{\ell-1}^{b}(t)+\gamma_{\ell+1}^{b}(t)\right]\\
+ & \frac{J}{2}\delta_{s,b}\left[\gamma_{\ell-1}^{a}(t)+\gamma_{\ell+1}^{a}(t)\right]\\
+ & \frac{\Delta}{2}\delta_{T}(t)\delta_{s,a}\left[\gamma_{\ell-1}^{b}(t)-\gamma_{\ell+1}^{b}(t)\right]\\
+ & \frac{\Delta}{2}\delta_{T}(t)\delta_{s,b}\left[\gamma_{\ell-1}^{a}(t)-\gamma_{\ell+1}^{a}(t)\right]
\end{alignat*}
for $s=a,b$, or separately
\begin{alignat}{1}
\frac{d\gamma_{\ell}^{a}}{dt}= & -\mu\gamma_{\ell}^{b}-\frac{J}{2}(\gamma_{\ell-1}^{b}+\gamma_{\ell+1}^{b})\nonumber \\
& +\frac{\Delta}{2}\delta_{T}(t)(\gamma_{\ell-1}^{b}-\gamma_{\ell+1}^{b}),\label{eq:EOMa}\\
\frac{d\gamma_{\ell}^{b}}{dt}= & +\mu\gamma_{\ell}^{a}+\frac{J}{2}(\gamma_{\ell-1}^{a}+\gamma_{\ell+1}^{a})\nonumber \\
& +\frac{\Delta}{2}\delta_{T}(t)(\gamma_{\ell-1}^{a}-\gamma_{\ell+1}^{a}).\label{eq:EOMb}
\end{alignat}
The Floquet operator of the system in the Majorana basis $(\gamma_{1}^{a},\gamma_{2}^{a},...,\gamma_{L}^{a},\gamma_{1}^{b},\gamma_{2}^{b},...,\gamma_{L}^{b})$
is then obtained by integrating Eqs.~(\ref{eq:EOMa}) and (\ref{eq:EOMb})
over a driving period, i.e., from $t=mT+0^{-}$ to $(m+1)T+0^{-}$ with $m\in{\mathbb Z}$.
The resulting Floquet matrix can be cast into the form
\begin{equation}
{\cal U}=e^{-i{\cal H}_{2}}e^{-i{\cal H}_{1}},
\end{equation}
where ${\cal H}_{1}$ and ${\cal H}_{2}$ are both $2L\times2L$ block off-diagonal matrices
in the Majorana basis, given by
\begin{equation}
{\cal H}_{1}=\frac{\Delta}{2i}\sigma_{x}\otimes\begin{pmatrix}0 & 1 & 0 & 0 & \cdots & 0\\
-1 & 0 & 1 & 0 & \cdots & 0\\
0 & -1 & \ddots & \ddots & \ddots & \vdots\\
0 & 0 & \ddots & \ddots & 1 & 0\\
\vdots & \vdots & \ddots & -1 & 0 & 1\\
0 & 0 & \cdots & 0 & -1 & 0
\end{pmatrix}_{L\times L},
\end{equation}
\begin{equation}
{\cal H}_{2}=\sigma_{y}\otimes\begin{pmatrix}\mu & J/2 & 0 & 0 & \cdots & 0\\
J/2 & \mu & J/2 & 0 & \cdots & 0\\
0 & J/2 & \ddots & \ddots & \ddots & \vdots\\
0 & 0 & \ddots & \ddots & J/2 & 0\\
\vdots & \vdots & \ddots & J/2 & \mu & J/2\\
0 & 0 & \cdots & 0 & J/2 & \mu
\end{pmatrix}_{L\times L},
\end{equation}
where $\sigma_{x,y}$ are Pauli matrices in their usual representations.
The quasienergy spectrum and Floquet eigenstates can then be obtained
by diagonalizing the Floquet operator ${\cal U}$ in the Majorana
representation. 

\section{Floquet operator in BdG representation\label{subsec:UBdG}}

Instead of Majorana basis, it is also possible to express the Floquet
operator of the periodically kicked NHKC described by Eq.~(\ref{eq:HtOBC})
in fermionic basis. This is achieved by performing the following Bogoliubov
transformation~\cite{BdGBook}:
\begin{equation}
c_{n}=\sum_{j=1}^{L}(u_{jn}f_{j}+v_{jn}f_{j}^{\dagger}),\label{eq:CnInBdG}
\end{equation}
where $f_{j}^{\dagger}$ and $f_{j}$ are creation and annihilation
operators of normal fermions, satisfying the anticommutation relations
\begin{equation}
\{f_{i},f_{j}\}=\{f_{i}^{\dagger},f_{j}^{\dagger}\}=0,\qquad\{f_{i},f_{j}^{\dagger}\}=\delta_{ij}.
\end{equation}
The coefficients $u_{jn},v_{jn}$ are chosen to be real, satisfying
the normalization condition
\begin{equation}
\sum_{j=1}^{L}(u_{jn}^{2}+v_{jn}^{2})=1.
\end{equation}
The Hamiltonian $H(t)$ of the periodically kicked NHKC is assumed to be diagonalized in the
basis $\{f_{1},...,f_{L},f_{1}^{\dagger},...,f_{L}^{\dagger}\}$,
such that it can be expressed as
\begin{equation}
H(t)=\sum_{\ell}E_{\ell}f_{\ell}^{\dagger}f_{\ell}.\label{eq:HtInBdG}
\end{equation}
Using the relation (\ref{eq:ComABC}), it is straightforward to show
that 
\begin{equation}
[H(t),f_{j}]=-E_{j}f_{j},\qquad[H(t),f_{j}^{\dagger}]=E_{j}f_{j}^{\dagger},
\end{equation}
and
\begin{equation}
[H(t),c_{n}]=-\sum_{j}E_{j}(u_{jn}f_{j}-v_{jn}f_{j}^{\dagger}).\label{eq:HtCn1}
\end{equation}
In the meantime, we can also compute the commutator $[H(t),c_{n}]$
directly with the $H(t)$ in Eq.~(\ref{eq:HtOBC}) and $c_{n}$ in
Eq.~(\ref{eq:CnInBdG}). The result is
\begin{alignat}{1}
& [H(t),c_{n}]=-\sum_{j}\mu(u_{jn}f_{j}+v_{jn}f_{j}^{\dagger})\label{eq:HtCn2}\\
- & \sum_{j}\frac{J}{2}[(u_{jn-1}+u_{jn+1})f_{j}+(v_{jn-1}+v_{jn+1})f_{j}^{\dagger}]\nonumber \\
- & \sum_{j}\frac{\Delta}{2}\delta_{T}(t)[(v_{jn-1}-v_{jn+1})f_{j}+(u_{jn-1}-u_{jn+1})f_{j}^{\dagger}].\nonumber 
\end{alignat}
Comparing Eqs.~(\ref{eq:HtCn1}) and (\ref{eq:HtCn2}), we obtain
the following BdG self-consistent equations (after dropping the redundant
index $j$):
\begin{alignat}{1}
Eu_{n}= & +\mu u_{n}+\frac{J}{2}(u_{n-1}+u_{n+1})+\frac{\Delta}{2}\delta_{T}(t)(v_{n-1}-v_{n+1})\label{eq:Eun}\\
Ev_{n}= & -\mu v_{n}-\frac{J}{2}(v_{n-1}+v_{n+1})-\frac{\Delta}{2}\delta_{T}(t)(u_{n-1}-u_{n+1})\label{eq:Evn}
\end{alignat}
The Floquet operator of the system in the BdG basis can then be obtained
by integrating Eqs.~(\ref{eq:Eun}) and (\ref{eq:Evn}) over a driving
period, i.e., from $t=mT+0^{-}$ to $(m+1)T+0^{-}$ with $m\in{\mathbb Z}$, leading to the
Floquet matrix:
\begin{equation}
\mathsf{U}=e^{-i\mathsf{H}_{2}}e^{-i\mathsf{H}_{1}}.
\end{equation}
Here $\mathsf{H}_{1}$ is a block off-diagonal matrix of the form:
\begin{equation}
\mathsf{H}_{1}=\frac{\Delta}{2i}\sigma_{y}\otimes\begin{pmatrix}0 & 1 & 0 & 0 & \cdots & 0\\
-1 & 0 & 1 & 0 & \cdots & 0\\
0 & -1 & \ddots & \ddots & \ddots & \vdots\\
0 & 0 & \ddots & \ddots & 1 & 0\\
\vdots & \vdots & \ddots & -1 & 0 & 1\\
0 & 0 & \cdots & 0 & -1 & 0
\end{pmatrix}_{L\times L},
\end{equation}
and $\mathsf{H}_{2}$ is a $2L\times2L$ tridiagonal matrix of the form:
\begin{equation}
\mathsf{H}_{2}=\sigma_{z}\otimes\begin{pmatrix}\mu & J/2 & 0 & 0 & \cdots & 0\\
J/2 & \mu & J/2 & 0 & \cdots & 0\\
0 & J/2 & \ddots & \ddots & \ddots & \vdots\\
0 & 0 & \ddots & \ddots & J/2 & 0\\
\vdots & \vdots & \ddots & J/2 & \mu & J/2\\
0 & 0 & \cdots & 0 & J/2 & \mu
\end{pmatrix}_{L\times L},
\end{equation}
with $\sigma_{y,z}$ being Pauli matrices in their usual representations.
So the quasienergy spectrum and Floquet eigenstates of the periodically kicked NHKC under OBC
can also be obtained by diagonalizing the Floquet operator $\mathsf{U}$
in the BdG basis. In fact, the Floquet operators ${\cal U}$ in Majorana
basis and $\mathsf{U}$ in BdG basis can be simply mapped into each
other by swapping the Pauli matrices $\sigma_{x}\leftrightarrow\sigma_{y}$
and $\sigma_{y}\leftrightarrow\sigma_{z}$, which does not change
the quasienergy spectrum as expected.

\section{Finite size effect\label{subsec:FSE}}
\begin{figure}
	\begin{centering}
		\includegraphics[scale=0.475]{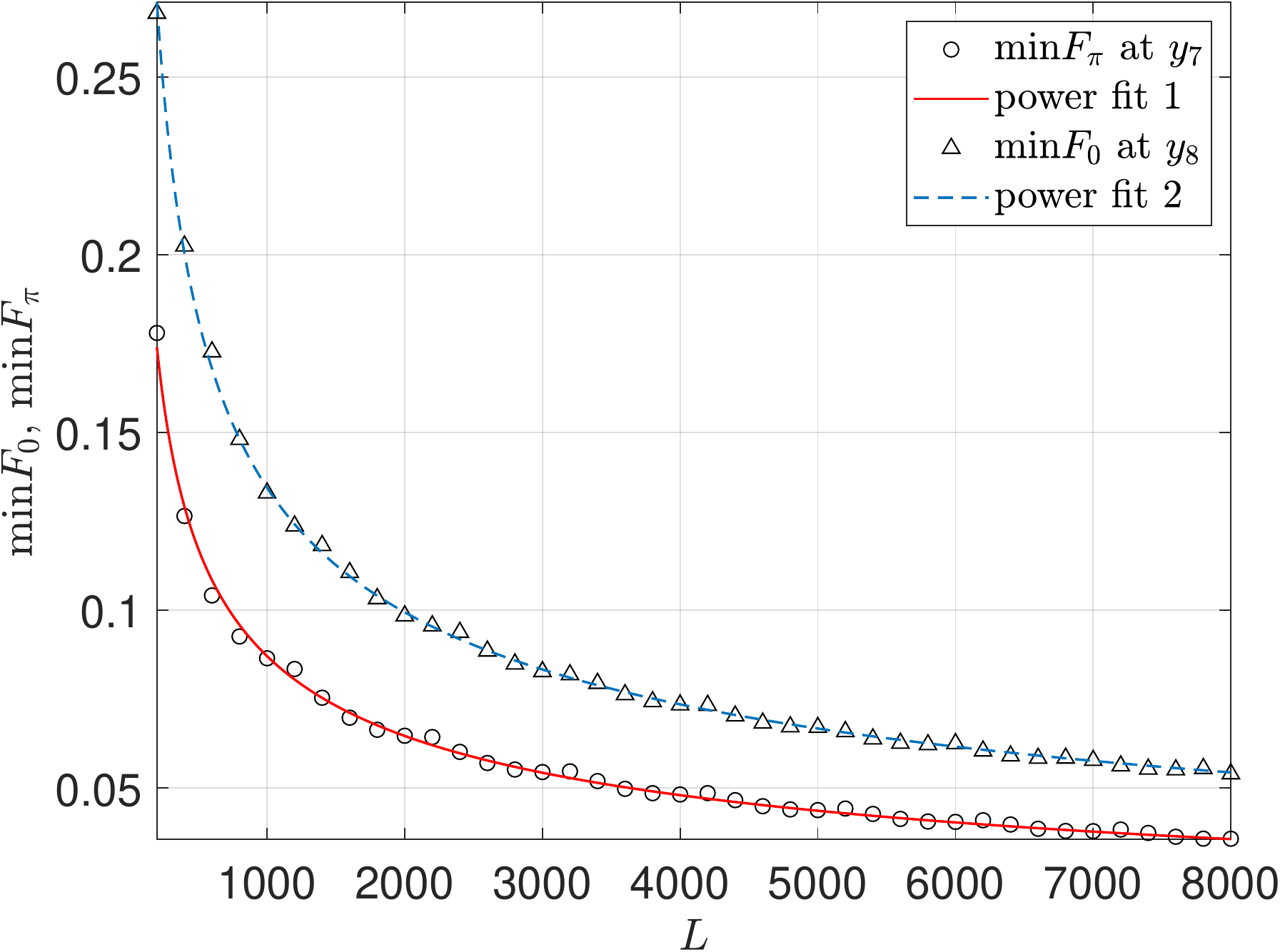}
		\par\end{centering}
	\caption{The minimum of gap functions $F_\pi$ (circles) and $F_0$ (triangles) versus the system size $L$ at the phase transition points $y_7$ and $y_8$ in Fig.~\ref{fig:SpectrumOBC1} of the main text, respectively. The red solid and blue dashed lines are the power fittings ($y=aL^b$) of the corresponding minimum gap functions min$F_\pi$ and min$F_0$. For the power fit 1 (power fit 2), we have $a=1.7,b=-0.4302$ ($a=2.725,b=-0.4356$) with $95\%$ confidence bounds. The fact that $b<0$ for both curves indicates $\lim_{L\rightarrow\infty} \min F_\pi\rightarrow0$ and $\lim_{L\rightarrow\infty} \min F_0\rightarrow0$ at $y_7$ and $y_8$ in Fig.~\ref{fig:SpectrumOBC1}, respectively.}\label{fig:F0PScaling}
\end{figure}
In Fig.~\ref{fig:SpectrumOBC1}, we could observe deviations between the theoretically predicted phase transition points (i.e., $\mu_i=y_1,...,y_8$) and the numerical behaviors of gap functions $F_0$ and $F_\pi$ around these points. The deviations are most obvious at $\mu_i=y_7$ and $y_8$. In this appendix, we show that these deviations are due to finite-size effects, which tend to vanish in the thermodynamic limit $L\rightarrow\infty$.

In Fig.~\ref{fig:F0PScaling}, we plot the minimum of $F_\pi$ (circles) and $F_0$ (triangles) versus the size of the chain $L$ right at the phase transition points $\mu_i=y_7$ and $\mu_i=y_8$, respectively. The other system parameters are chosen to be the same as in Fig.~\ref{fig:SpectrumOBC1}. We see that both $F_\pi$ and $F_0$ are indeed decreasing with the increase of $L$. Furthermore, we performed power fittings to the data points of $F_\pi$ and $F_0$, and found that both of them scale with the size of the lattice $L$ in the functional form $y=aL^b$, with $b<0$. Therefore, in the limit $L\rightarrow\infty$, we will have $F_\pi\rightarrow0$ and $F_0\rightarrow0$ at $\mu_i=y_7$ and $\mu_i=y_8$, which demonstrates that the deviations observed in Fig.~\ref{fig:SpectrumOBC1} are finite-size effects.

In the meantime, we notice that when $\mu_i$ is large, it will dominate the behaviors of gap functions $F_0,F_\pi$ versus $\mu_i$ around the phase transition points, and therefore control how the corresponding gaps approach zero. Our numerical calculations suggest that close to a transition point, a larger $\mu_i$ would introduce stronger repulsions between Majorana zero or $\pi$ modes, pushing them away from zero or $\pi$ quasienergies faster when approaching transition points at large values of $\mu_i$ (i.e. $y_7,y_8$ in Fig.~\ref{fig:SpectrumOBC1}).



\begin{thebibliography}{99}
	
	\bibitem{NHBook1} N. Moiseyev, \textit{Non-Hermitian Quantum Mechanics} (Cambridge University Press, Cambridge, England, 2011).
	
	\bibitem{NHBook2} C. M. Bender, \textit{PT Symmetry In Quantum and Classical Physics} (World Scientific, Singapore, 2019).
	
	\bibitem{NHTPReview1} Z. Gong, Y. Ashida, K. Kawabata, K. Takasan, S. Higashikawa, M. Ueda, Phys. Rev. X {\bf 8}, 031079 (2018).
	
	\bibitem{NHTPReview2} H. Shen, B. Zhen, and L. Fu, Phys. Rev. Lett. {\bf 120}, 146402 (2018).
	
	\bibitem{NHTPReview3} V. M. M. Alvarez, J. E. B. Vargas, M. Berdakin, and L. E. F. Foa Torres, Eur. Phys. J. Special Topics {\bf 227}, 1295 (2018).
	
	\bibitem{NHTPReview4} A. Ghatak, and T. Das, J. Phys.: Condens. Matter {\bf 31}, 263001 (2019).
	
	\bibitem{NHTPReview5} L. E. F. Foa Torres, J. Phys.: Mater. {\bf 3}, 014002 (2020).
	
	\bibitem{NHTPReview6} H. Zhao and L. Feng, National Science Review {\bf 5}, 183-199 (2018).
	
	\bibitem{NHTPReview7} H. Cao and J. Wiersig, Rev. Mod. Phys. {\bf 87}, 61 (2015).
	
	\bibitem{NHTPReview8} S. Longhi, EPL {\bf 120}, 64001 (2017).
	
	\bibitem{NHTPReview9} S. K. \"Ozdemir, S. Rotter, F. Nori, and L. Yang, Nature Materials {\bf 18}, 783-798 (2019).
	
	\bibitem{NHTPReview10} M. Miri and A. Alu, Science {\bf 363}, eaar7709 (2019).
	
	\bibitem{NHTPOS1} V. V. Konotop, J. Yang, and D. A. Zezyulin, Rev. Mod. Phys. {\bf 88}, 035002 (2016).
	
	\bibitem{NHTPOS2} L. Feng, R. El-Ganainy, and L. Ge, Nat. Photon. {\bf 11}, 752 (2017).
	
	\bibitem{NHTPOS3} R. El-Ganainy, K. G. Makris, M. Khajavikhan, Z. H. Musslimani, S. Rotter, and D. N. Christodoulides, Nat. Phys. {\bf 14}, 11 (2018).
	
	\bibitem{NHTPOS4} T. Yoshida, K. Kudo, and Y. Hatsugai, Sci. Rep. {\bf 9}, 16895 (2019).
	
	\bibitem{NHTPBOSE1} R. Barnett, Phys. Rev. A {\bf 88}, 063631 (2013).
	
	\bibitem{NHTPBOSE2} B. Galilo, D. K. K. Lee, and R. Barnett, Phys. Rev. Lett. {\bf 115}, 245302 (2015).
	
	\bibitem{NHTPBOSE3} G. Engelhardt and T. Brandes, Phys. Rev. A {\bf 91}, 053621 (2015).
	
	\bibitem{NHTPBOSE4} G. Engelhardt, M. Benito, G. Platero, and T. Brandes, Phys. Rev. Lett. {\bf 117}, 045302 (2016).
	
	\bibitem{NHTPBOSE5} V. Peano, M. Houde, F. Marquardt, and A. A. Clerk, Phys. Rev. X {\bf 6}, 041026 (2016).
	
	\bibitem{NHTPBOSE6} S. Lieu, Phys. Rev. B {\bf 98}, 115135 (2018).
	
	\bibitem{NHTPBOSE7} A. McDonald, T. Pereg-Barnea, and A. A. Clerk, Phys. Rev. X {\bf 8}, 041031 (2018).
	
	\bibitem{NHTPFERMI1} M. Papaj, H. Isobe, and L. Fu, Phys. Rev. B {\bf 99}, 201107(R) (2019).
	
	\bibitem{NHTPFERMI2} H. Shen and L. Fu,	Phys. Rev. Lett. {\bf 121},	026403 (2018).
	
	\bibitem{NHTPFERMI3} K. Moors, A. A. Zyuzin, A. Y. Zyuzin, R. P. Tiwari, and T. L. Schmidt, Phys. Rev. B {\bf 99}, 041116(R) (2019).
	
	\bibitem{NHTPFERMI4} T. Yoshida, R. Peters, N. Kawakami, and Y. Hatsugai, Phys. Rev. B {\bf 99}, 121101(R) (2019).
	
	\bibitem{NHTPFERMI5} T. M. Philip, M. R. Hirsbrunner, and M. J. Gilbert, Phys. Rev. B {\bf 98}, 155430 (2018).
	
	\bibitem{NHTPFERMI6}  Y. Chen and H. Zhai, Phys. Rev. B {\bf 98}, 245130 (2018).
	
	\bibitem{NHTPFERMI7} E. J. Bergholtz and J. C. Budich, Phys. Rev. Research {\bf 1}, 012003(R) (2019).
	
	\bibitem{NHTPMETA1} C. L. Kane and T. C. Lubensky, Nat. Phys. {\bf 10}, 39 (2014).
	
	\bibitem{NHTPMETA2} K. Roychowdhury, D. Z. Rocklin, and M. J. Lawler, Phys. Rev. Lett. {\bf 121}, 177201 (2018).
	
	\bibitem{NHTPMETA3} K. Roychowdhury and M. J. Lawler, Phys. Rev. B {\bf 98}, 094432 (2018).
	
	\bibitem{NHTPInvis1} Z. Lin, H. Ramezani, T. Eichelkraut, T. Kottos, H. Cao, and D. N. Christodoulides, Phys. Rev. Lett. {\bf 106}, 213901 (2011).
	
	\bibitem{NHTPInvis2} A. Regensburger, C. Bersch, M.-A. Miri, G. Onishchukov, D. N. Christodoulides, and U. Peschel, Nature (London) {\bf 488}, 167 (2012).
	
	\bibitem{NHTPInvis3} L. Feng, Y.-L. Xu, W. S. Fegadolli, M.-H. Lu, J. E. B. Oliveira, V. R. Almeida, Y.-F. Chen, and A. Scherer,	Nat. Mater. {\bf 12}, 108 (2013).
	
	\bibitem{NHTPInvis4} B. Peng, S. K. \"Ozdemir, F. Lei, F. Monifi, M. Gianfreda,	G. L. Long, S. Fan, F. Nori, C. M. Bender, and L. Yang,	Nat. Phys. {\bf 10}, 394 (2014).
	
	\bibitem{NHTPSens1} J. Wiersig, Phys. Rev. Lett. {\bf 112}, 203901 (2014).
	
	\bibitem{NHTPSens2} Z.-P. Liu, J. Zhang, ?. K. Özdemir, B. Peng, H. Jing, X.-Y. L\"u, C.-W. Li, L. Yang, F. Nori, and Y. x. Liu, Phys. Rev. Lett. {\bf 117}, 110802 (2016).
	
	\bibitem{NHTPSens3} H.-K. Lau and A. A. Clerk, Nat. Commun. {\bf 9}, 4320 (2018).
	
	\bibitem{NHTPSens4} H. Hodaei, A. U. Hassan, S. Wittek, H. Garcia-Gracia, R. El-Ganainy, D. N. Christodoulides, and M. Khajavikhan,	Nature (London) {\bf 548}, 187 (2017).
	
	\bibitem{NHTPSens5}  W. Chen, S. K. \"Ozdemir, G. Zhao, J. Wiersig, and L. Yang, Nature (London) {\bf 548}, 192 (2017).
	
	\bibitem{TPLaser1} G. Harari, M. A. Bandres, Y. Lumer, M. C. Rechtsman, Y. D. Chong, M. Khajavikhan, D. N. Christodoulides, and M. Segev, Science {\bf 359}, eaar4003 (2018).
	
	\bibitem{TPLaser2} M. A. Bandres, S. Wittek, G. Harari, M. Parto, J. Ren, M. Segev, D. N. Christodoulides, and M. Khajavikhan, Science {\bf 359}, eaar4005 (2018).
	
	\bibitem{TPLaser3} Y. V. Kartashov and D. V. Skryabin, Phys. Rev. Lett. {\bf 122}, 083902 (2019).
	
	\bibitem{TPLaser4} S. Longhi, Annalen der Physik (Berlin) {\bf 530}, 1800023 (2018).
	
	\bibitem{TPLaser5} Y. D. Chong, L. Ge, and A. D. Stone, Phys. Rev. Lett. {\bf 106}, 093902 (2011).
	
	\bibitem{TPLaser6} M. Parto, S. Wittek, H. Hodaei, G. Harari, M. A. Bandres, J. Ren, and M. C. Rechtsman, Phys. Rev. Lett. {\bf 120}, 113901 (2018).
	
	\bibitem{TPLaser7} G. Oktay, M. Sarisaman, and M. Tas, arXiv:1905.05799.
	
	\bibitem{TPLaser8} I. Amelio and I. Carusotto, arXiv:1911.10437.
	
	\bibitem{Class1} C. Liu, H. Jiang, and S. Chen, Phys. Rev. B {\bf 99}, 125103 (2019).
	
	\bibitem{Class2} K. Kawabata, S. Higashikawa, Z. Gong, Y. Ashida and M. Ueda, Nat. Commun. {\bf 10}, 297 (2019).
	
	\bibitem{Class3} Z. Ge, Y. Zhang, T. Liu, S. Li, H. Fan, and F. Nori, Phys. Rev. B {\bf 100}, 054105 (2019).
	
	\bibitem{Class4} K. Kawabata, T. Bessho, and M. Sato, Phys. Rev. Lett. {\bf 123}, 066405 (2019).
	
	\bibitem{Class5} L. Li, C. H. Lee, and J. Gong, Phys. Rev. B {\bf 100}, 075403 (2019).
	
	\bibitem{Class6} H. Zhou and J. Y. Lee, Phys. Rev. B {\bf 99} 235112 (2019).
	
	\bibitem{Class7} K. Kawabata, K. Shiozaki, M. Ueda, and M. Sato, Phys. Rev. X {\bf 9}, 041015 (2019).
	
	\bibitem{Class8} J. Y. Lee, J. Ahn, H. Zhou, and A. Vishwanath, Phys. Rev. Lett. {\bf 123}, 206404 (2019).
	
	\bibitem{Class9} W. B. Rui, Y. X. Zhao, and A. P. Schnyder, Phys. Rev. B {\bf 99}, 241110 (2019).
	
	\bibitem{Class10} C. Liu and S. Chen, Phys. Rev. B {\bf 100}, 144106 (2019).
	
	\bibitem{Class11} T. Yoshida and Y. Hatsugai, Phys. Rev. B {\bf 100}, 054109 (2019).
	
	\bibitem{Class12} Z. Li and R. S. K. Mong, arXiv:1911.02697.
	
	\bibitem{Class13} W. Xi, Z. Zhang, Z. Gu, and W. Chen, arXiv:1911.01590.
	
	\bibitem{BBC0} S. Yao and Z. Wang, Phys. Rev. Lett. {\bf 121}, 086803 (2018).
	
	\bibitem{BBC1} F. K. Kunst, E. Edvardsson, J. C. Budich, and E. J. Bergholtz, Phys. Rev. Lett. {\bf 121}, 026808 (2018).
	
	\bibitem{BBC2} D. S. Borgnia, A. J. Kruchkov, and R. Slager, arXiv:1902.07217.
	
	\bibitem{BBC3} E. Edvardsson, F. K. Kunst, and E. J. Bergholtz, Phys. Rev. B {\bf 99}, 081302 (2019).
	
	\bibitem{BBC4} L. Jin and Z. Song, Phys. Rev. B {\bf 99}, 081103 (2019).
	
	\bibitem{BBC5} H. Wang, J. Ruan, and H. Zhang, Phys. Rev. B {\bf 99}, 075130 (2019).
	
	\bibitem{BBC6} H. Zirnstein, G. Refael, and B. Rosenow, arXiv:1901.11241.
	
	\bibitem{BBC7} L. Herviou, J. H. Bardarson, and N. Regnault, Phys. Rev. A {\bf 99}, 052118 (2019).
	
	\bibitem{BBC8} K. Imura and Y. Takane, Phys. Rev. B {\bf 100}, 165430 (2019).
	
	\bibitem{BBC9} F. K. Kunst and V. Dwivedi, Phys. Rev. B {\bf 99}, 245116 (2019).
	
	\bibitem{BBC10} F. Song, S. Yao, and Z. Wang, Phys. Rev. Lett. {\bf 123}, 246801 (2019).
	
	\bibitem{BBC11} Y. Xiong, J. Phys. Commun. {\bf 2}, 035043 (2018).
	
	\bibitem{BBC12} C. H. Lee and R. Thomale, Phys. Rev. B {\bf 99}, 201103(R) (2019).
	
	\bibitem{Exp51} A. Ghatak, M. Brandenbourger, J. v. Wezel, and C. Coulais, arXiv:1907.11619.
	
	\bibitem{Exp52} W. Zhu, X. Fang, D. Li, Y. Sun, Y. Li, Y. Jing, and H. Chen, Phys. Rev. Lett. {\bf 121}, 124501 (2018).
	
	\bibitem{Exp11} J. M. Zeuner, M. C. Rechtsman, Y. Plotnik, Y. Lumer, S. Nolte, M. S. Rudner, M. Segev, and A. Szameit, Phys. Rev. Lett. {\bf 115}, 040402 (2015).
	
	\bibitem{Exp12} M. Hafezi, E. Demler, M. Lukin, J. Taylor, Nat. Phys. {\bf 7}, 907 (2011).
	
	\bibitem{Exp13} P. Peng, W. Cao, C. Shen, W. Qu, J. Wen, L. Jiang and Y. Xiao, Nat. Phys. {\bf 12}, 1139-1145 (2016).
	
	\bibitem{Exp21} S. Weimann, M. Kremer, Y. Plotnik, Y. Lumer, S. Nolte, K. G. Makris, M. Segev, M. C. Rechtsman, and A. Szameit, Nature Mater. {\bf 16}, 433 (2016).
	
	\bibitem{Exp22} L. Xiao, T. Deng, K. Wang, G. Zhu, Z. Wang, W. Yi, and P. Xue, arXiv:1907.12566.
	
	\bibitem{Exp23} K. Wang, X. Qiu, L. Xiao, X. Zhan, Z. Bian, B. C. Sanders, W. Yi, and P. Xue, Nature Commun. {\bf 10}, 2293 (2019).
	
	\bibitem{Exp41} C. Poli, M.	Bellec, U. Kuhl, F. Mortessagne, and H. Schomerus, Nature Commun. {\bf 6}, 6710 (2015).
	
	\bibitem{Exp42} H. Xu, D. Mason, L. Jiang, J.G.E. Harris, Nature {\bf 537}, 80 (2016).
	
	\bibitem{Exp31} T. Helbig, T. Hofmann, S. Imhof, M. Abdelghany, T. Kiessling, L. W. Molenkamp, C. H. Lee, A. Szameit, M. Greiter and R. Thomale, arXiv:1907.11562.
	
	\bibitem{Exp32} T. Hofmann, T. Helbig, F. Schindler, N. Salgo, M. Brzezinska, M. Greiter, T. Kiessling, D. Wolf, A. Vollhardt, A. Kabasi, C. H. Lee, A. Bilusic, R. Thomale, and T. Neupert, arXiv:1908.02759.
	
	\bibitem{ZhouPRB2019} L. Zhou, Phys. Rev. B {\bf 100}, 184314 (2019).
	
	\bibitem{ZhouPRB2018} L. Zhou and J. Gong, Phys. Rev. B {\bf 98} 205417 (2018).
	
	\bibitem{ZhouPRA2019} L. Zhou and J. Pan, Phys. Rev. A {\bf 100}, 053608 (2019).
	
	\bibitem{YuceEPJD2015} C. Yuce, Eur. Phys. J. D {\bf 69}, 184 (2015).
	
	\bibitem{FehskePRL2019} B. H\"ockendorf, A. Alvermann, and H. Fehske, Phys. Rev. Lett. {\bf 123}, 190403 (2019).
	
	\bibitem{LiPRB2019} M. Li, X. Ni, M. Weiner, A. Alu, and A. B. Khanikaev, Phys. Rev. B {\bf 100}, 045423 (2019).
	
	\bibitem{NHSkin12} X. Zhang and J. Gong, Phys. Rev. B {\bf 101}, 045415 (2020).
	
	\bibitem{NHPump1} B. H\"ockendorf, A. Alvermann, and Holger Fehske, arXiv:1911.11413.
	
	
	
	\bibitem{MajoranaRev3} S. D. Sarma, M. Freedman and C. Nayak, npj Quantum Information {\bf 1}, 15001 (2015).
	
	\bibitem{MajoranaRev4} J. Alicea, Rep. Prog. Phys. {\bf 75}, 076501 (2012).
	
	\bibitem{MajoranaRev5} M. Leijnse and K. Flensberg, Semicond. Sci. Technol. {\bf 27} 124003 (2012).
	
	\bibitem{MajoranaRev6} C. W. J. Beenakker, Annu. Rev. Condens. Matter Phys. {\bf 4}, 113-136 (2013).
	
	\bibitem{MajoranaRev7} C. W. J. Beenakker, Rev. Mod. Phys. {\bf 87}, 1037 (2015).
	
	\bibitem{MajoranaRev8} M. Sato and S. Fujimoto, J. Phys. Soc. Jpn. {\bf 85}, 072001 (2016).
	
	\bibitem{MajoranaRev9} A. Haim and Y. Oreg, Phys. Rep. {\bf 825}, 1-48 (2019).
	
	\bibitem{NHMajorana0} M. T. van Caspel, S. E. T. Arze, and I. P. Castillo, SciPost Phys. {\bf 6}, 026 (2019).
	
	\bibitem{NHMajorana1} S. Lieu, Phys. Rev. B {\bf 100}, 085110 (2019).
	
	\bibitem{NHMajorana2} K. Kawabata, Y. Ashida, H. Katsura, and M. Ueda, Phys. Rev. B {\bf 98}, 085116 (2018).
	
	\bibitem{NHMajorana3} N. Okuma and M. Sato, Phys. Rev. Lett. {\bf 123}, 097701 (2019).
	
	\bibitem{NHMajorana4} C. Li, X. Z. Zhang, G. Zhang, and Z. Song, Phys. Rev. B {\bf 97}, 115436 (2018).
	
	\bibitem{NHMajorana5} N. Shibata and H. Katsura, Phys. Rev. B {\bf 99}, 174303 (2019).
	
	\bibitem{NHMajorana6} M. Ezawa, Phys. Rev. B {\bf 100}, 045407 (2019).
	
	\bibitem{NHMajorana7} C. Yuce, Phys. Rev. A {\bf 93}, 062130 (2016).
	
	\bibitem{NHMajorana8} Q.-B. Zeng, B. Zhu, S. Chen, L. You, and R. L\"u, Phys. Rev. A {\bf 94}, 022119 (2016).
	
	\bibitem{NHMajorana9} J. Avila, F. Penaranda, E. Prada, P. San-Jose, and R. Aguado, Communications Physics {\bf 2}, 133 (2019).
	
	\bibitem{NHMajorana11} M. Klett, H. Cartarius, D. Dast, J. Main, and G. Wunner, Phys. Rev. A {\bf 95}, 053626 (2017).
	
	\bibitem{NHMajorana12} X. Wang, T. Liu, Y. Xiong, and P. Tong, Phys. Rev. A {\bf 92}, 012116 (2015).
	
	\bibitem{NHMajorana13} A. A. Zyuzin and P. Simon, Phys. Rev. B {\bf 99}, 165145 (2019).
	
	\bibitem{NHMajorana14} P. San-Jose, J. Cayao, E. Prada, and R. Aguado, Sci. Rep. {\bf 6}, 21427 (2015).
	
	\bibitem{FloMaj0} L. Jiang, T. Kitagawa, J. Alicea, A. R. Akhmerov, D. Pekker, G. Refael, J. I. Cirac, E. Demler, M. D. Lukin, and P. Zoller, Phys. Rev. Lett. {\bf 106}, 220402 (2011).
	
	\bibitem{FloMaj1} Q.-J. Tong, J.-H. An, J. Gong, H.-G. Luo, and C. H. Oh, Phys. Rev. B \textbf{87}, 201109 (2013).
	
	\bibitem{FloMaj7} D. T. Liu, J. Shabani, and A. Mitra, Phys. Rev. B {\bf 99}, 094303 (2019).
	
	\bibitem{FloMaj8} D. E. Liu, A. Levchenko, and H. U. Baranger, Phys. Rev. Lett. {\bf 111}, 047002 (2013).
	
	\bibitem{FloMaj3} M. Thakurathi, A. A. Patel, D. Sen, and A. Dutta, Phys. Rev. B {\bf 88}, 155133 (2013).
	
	\bibitem{FloMaj4} P. Molignini, W. Chen, and and R. Chitra, Phys. Rev. B {\bf 98}, 125129 (2018).
	
	\bibitem{FloMaj5} P. Molignini, E. van Nieuwenburg, and R. Chitra, Phys. Rev. B {\bf 96}, 125144 (2017).
	
	\bibitem{FloMaj6} Y. Li, A. Kundu, F. Zhong, and B. Seradjeh, Phys. Rev. B {\bf 90}, 121401(R) (2014).
	
	\bibitem{FloMaj2} H. H. Yap, L. Zhou, C. H. Lee, J. Gong, Phys. Rev. B {\bf 97}, 165142 (2018).
	
	\bibitem{FloCompt1} R. W. Bomantara and J. Gong, Phys. Rev. Lett. {\bf 120}, 230405 (2018).
	
	\bibitem{FloCompt2} R. W. Bomantara and J. Gong, Phys. Rev. B {\bf 98}, 165421 (2018).
	
	\bibitem{FloCompt3} R. W. Bomantara and J. Gong, arXiv:1904.03161.
	
	\bibitem{FloCompt4} B. Bauer, T. Pereg-Barnea, T. Karzig, M.-T. Rieder, G. Refael, E. Berg, and Y. Oreg, Phys. Rev. B {\bf 100}, 041102(R) (2019).
	
	\bibitem{KC} A. Y. Kitaev, Sov. Phys. Usp. {\bf 44}, 131 (2001).
	
	\bibitem{Book3} F. Franchini, \textit{An Introduction to Integrable	Techniques for One-Dimensional Quantum Systems} (Springer, Cham, Switzerland, 2017).
	
	\bibitem{Tenfold} S. Ryu, A. P Schnyder, A. Furusaki, and Andreas W. W. Ludwig,  New J. Phys. {\bf 12} 065010 (2010).
	
	\bibitem{CooperRMPCold} N. R. Cooper, J. Dalibard, and I. B. Spielman, Rev. Mod. Phys. {\bf 91}, 015005 (2019).
	
	\bibitem{KCInt} L. Fidkowski and A. Kitaev, Phys. Rev. B {\bf 81}, 134509 (2010); L. Fidkowski and A. Kitaev, Phys. Rev. B {\bf 83}, 075103 (2011).
	
	\bibitem{NHKCExp1} T. E. Lee and C.-K. Chan, Phys. Rev. X {\bf 4}, 041001 (2014); T. E. Lee, F. Reiter, and N. Moiseyev, Phys. Rev. Lett. {\bf 113}, 250401 (2014).
	
	\bibitem{NHKCExp2} Y. Ashida, S. Furukawa, and M. Ueda, Phys. Rev. A {\bf 94}, 053615
	(2016); Nat. Commun. {\bf 8}, 15791 (2017).
	
	\bibitem{NHKCExp3} J. Li, A. K. Harter, J. Liu, L. de Melo, Y. N. Joglekar, and L. Luo, Nat. Commun. {\bf 10}, 855 (2019).
	
	\bibitem{AsbothSTF} J. K. Asb$\acute{{\rm o}}$th, Phys. Rev. B \textbf{86},	195414 (2012); J. K. Asb$\acute{{\rm o}}$th, and H. Obuse, Phys. Rev. B \textbf{88}, 121406 (2013).
	
	\bibitem{AntiTrap} Y. Xu, S. Wang, and L.-M. Duan, Phys. Rev. Lett. {\bf 118}, 045701 (2017).
	
	\bibitem{PairingExp1} A. Bhler, N. Lang, C. V. Kraus, G. Miller, S. D. Huber, and H. P. Bchler, Nat. Commun. {\bf 5}, 4504 (2014).
	
	\bibitem{SynSOC} C. Zhang, S. Tewari, R. M. Lutchyn, and S. Das Sarma, Phys. Rev. Lett. {\bf 101}, 160401 (2008).
	
	\bibitem{FeshbachRes} C. A. Regal, C. Ticknor, J. L. Bohn, and D. S. Jin, Phys. Rev. Lett. {\bf 90}, 053201 (2003).
	
	\bibitem{FeshbachRes2} F. A. van Abeelen and B. J. Verhaar, Phys. Rev. Lett. {\bf 83}, 1550 (1999).
	
	\bibitem{FeshbachRes3} D. M. Bauer, M. Lettner, C. Vo, G. Rempe, and S. D\"urr, Nat. Phys. {\bf 5}, 339-342 (2009).
	
	\bibitem{SynSOC2} Y.-J. Lin, K. Jim\'enez-Garc\'ia, and I. B. Spielman, Nature {\bf 471}, 83-86 (2011).
	
	\bibitem{DynamicWN} B. Zhu, Y. Ke, H. Zhong, and C. Lee, arXiv:1907.11348.
	
	\bibitem{DynamicST} C. Yi, L. Zhang, L. Zhang, R. Jiao, X. Cheng, Z. Wang, X. Xu, W. Sun, X. Liu, S. Chen, and J. Pan, Phys. Rev. Lett. {\bf 123}, 190603 (2019).
	
	\bibitem{MCDExp} D. Xie, W. Gou, T. Xiao,  B. Gadway, and B. Yan, npj Quantum Inf. {\bf 5}, 55 (2019).
	
	\bibitem{BdGBook} P. G. de Gennes, \textit{Superconductivity of Metals and Alloys} (CRC Press, New York, US, 1999).
	
\end{thebibliography}
\end{document}